\documentclass[transmag]{IEEEtran}
\usepackage[ruled]{algorithm2e}
\usepackage{amsfonts}
\usepackage{amssymb}
\usepackage{eurosym}
\usepackage{authblk}
\usepackage{cite}
\usepackage{graphicx}
\usepackage{amsmath}
\usepackage[labelformat=simple]{subcaption}
\usepackage{float}
\usepackage[T1]{fontenc}
\usepackage{slashbox}
\usepackage{amsthm}
\usepackage{mathrsfs}
\usepackage{lipsum}
\usepackage{mathtools}
\usepackage{multicol}
\usepackage{textcomp}
\usepackage{listings}
\usepackage{color}
\usepackage{amsmath}
\usepackage{eurosym}
\usepackage{mathrsfs}
\usepackage{hyperref}

\IEEEoverridecommandlockouts
\newcounter{mytempeqncnt}
\newtheorem{remark}{Remark}

\newtheorem{lemma}{Lemma}
\newtheorem{proposition}{Proposition}

\DeclareMathOperator{\Tr}{Tr}
\begin{document}

\title{A PHY Layer Security Analysis of a Hybrid High Throughput Satellite
with an Optical Feeder Link}
\author{Elmehdi Illi, \IEEEmembership{Member, IEEE}, Faissal El Bouanani, %
\IEEEmembership{Senior Member, IEEE}, Fouad Ayoub,
\IEEEmembership{Member,
IEEE}, \\
and Mohamed-Slim Alouini, \IEEEmembership{Fellow, IEEE} \thanks{%
E. Illi and F. El Bouanani are with ENSIAS College of Engineering, Mohammed
V University, Rabat, Morocco (e-mails: \{elmehdi.illi,
f.elbouanani\}@um5s.net.ma).} \thanks{%
F. Ayoub is with CRMEF, Kenitra, Morocco (e-mail: ayoub@crmefk.ma).} \thanks{
M.-S. Alouini is with Computer, Electrical, and Mathematical Sciences and
Engineering (CEMSE) Division, King Abdullah University of Science and
Technology (KAUST), Thuwal 23955-6900, Makkah Province, Saudi Arabia
(e-mail: slim.alouini@kaust.edu.sa).}}

\IEEEtitleabstractindextext{\begin{abstract}
Hybrid terrestrial-satellite (HTS) communication systems have gained a
tremendous amount of interest recently due to the high demand for global
high data rates. Conventional satellite communications operate in the
conventional Ku (12 GHz) and Ka (26.5-40 GHz) radio-frequency bands for
assessing the feeder link, between the ground gateway and the satellite.
Nevertheless, with the aim to provide hundreds of Mbps of throughput per
each user, free-space optical (FSO) feeder links have been proposed to
fulfill these high data rates requirements. In this paper, we investigate
the physical layer security performance for a hybrid very high throughput
satellite communication system with an {FSO} feeder link. In particular, the
satellite receives the incoming optical wave from an appropriate optical
ground station, carrying the data symbols of $N$ users through various
optical apertures and combines them using the selection combining technique.
Henceforth, the decoded and regenerated information signals of the $N$ users
are zero-forcing {(ZF)} precoded in order to cancel the interbeam
interference at the end-users. The communication is performed under the
presence of malicious eavesdroppers nodes at both hops. Statistical
properties of the signal-to-noise ratio of the legitimate and wiretap links
at each hop are derived, based on which the intercept probability metric is
evaluated. {The derived results show that above a certain number of optical
apertures, the secrecy level is not improved further. Also, the system's
secrecy is improved using ZF precoding compared to the no-precoding scenario
for some specific nodes' positions. All the derived analytical expressions
are validated through Monte Carlo simulations.}
\end{abstract}
}
\maketitle
\section{Introduction}

\IEEEPARstart{T}{hroughout} the last few years, satellite communication
(SatCom) has been a tremendously evolving segment of the wireless
communication industry, due to the increasing global demand on broadband
satellite communication links \cite{kaushal}. Interestingly, with the
arrival of the fifth-generation (5G) wireless cellular network, a variety of
satellite operators on the globe are developing broadband communications to
complement and compete with the terrestrial cellular networks \cite{ahmad}.
In this regard, multibeam SatCom has been widely advocated as an appropriate
way to assess very high-speed satellite-users link, in which a large number
of spot beams is used. Satellite links aim at providing high-speed
communications (order of hundreds of Mbps per user) to users in areas where
the traditional terrestrial networks offer very low quality of service \cite%
{katona}.

Among the critical challenges faced by the satellite communication industry
is the spectrum scarcity issue \cite{katona}, \cite{globalsip}. For
instance, with the high bandwidth requirement of the end-users, neither the
Ku-band (12 GHz) nor the Ka-band (26.5-40 GHz) seems to fulfill the hundreds
of Gbps aggregate user link throughput, due to the scarce frequency
resources in these bands \cite{qv1}. In addition to this, multiple ground
stations (gateways) are needed to feed the satellite to assess the desired
data rates while operating on such bands, which results in significant
energy consumption \cite{katona}.

To this end, free-space optics (FSO) technology has been broadly endorsed as
an effective solution for providing very high data rate links on
terrestrial-satellite communications \cite{fso1}. The overarching idea is to
carry on data from the optical ground station (OGS) in the form of conical
light beams using a powered laser device, operating either on the visible
(400-800 nm) or infrared (1500-1550 nm) spectrum, to a satellite, which
converts the optical signal to an electrical one, and serves the end-users
through radio-frequency (RF) spot beams \cite{globalsip}. From another
front, the use of optical bands does not require any regulation or license
fees as done in traditional Ku and Ka bands, where the International
Telecommunications Union (ITU) regulates their use. Also, besides its
immunity to interference and high security, FSO communication can provide a
data rate in the order of Tbps per optical beam, which renders it a viable
alternative solution to reach the desired high data rates \cite{fso1}. On
the other hand, optical communication is highly affected by atmospheric and
weather losses along the propagation path, known as turbulence. Pointing
error due to transmitter/receiver misalignment, as well as free-space
path-loss are two other limiting factors of FSO in outdoor communications
\cite{zedini}. Furthermore, another restricting phenomenon in
ground-to-satellite FSO\ links is cloud coverage. Indeed, optical
communication between the gateway and the satellite is blocked totally in
the presence of clouds \cite{globalsip}.

During the last decade, several works and researches have been conducted
onto the deployments of hybrid terrestrial-satellite (HTS) systems with an
optical feeder link. In \cite{globalsip}, the performance analysis of
optical feeder link with gateway diversity, in the presence of stochastic
cloud coverage model, is assessed. Furthermore, the authors in \cite{ahmad}
carried out a performance analysis of an HTS system with an optical feeder
link, with zero-forcing (ZF) precoding to remove the interbeam interference
(IBI), as well as amplification at the satellite are considered. In addition
to this, from an industrial perspective, several broadband optical
feeder-based HTS systems have been demonstrated in the last few years. For
instance, the first successful ground-to-satellite optical link was
performed between the earth and ETS-VI satellite in Konegi, Japan \cite%
{japansat}. Besides this, other HTS\ systems experiments demonstrated the
achievability of great throughput records, such as NASA's 622 Mbps Laser
communication experiments in 2014 \cite{nict1}, and DLR's Institute of
Communications and Navigation HTS experiments with a record data rate of
1.72 Tbit/s in 2016, and 13.16 Tbit/s in 2017 \cite{calvo2}. NICT is
planning in 2021 to launch a new test satellite with an aim to demonstrate a
10 Gbps speed on uplink and downlink of aggregate throughputs \cite{nictnews}%
. Moreover, in \cite{germany, calvo}, an overview of implementation,
performance, technological aspects, and users' quality of service on HTS
systems with an optical feeder is assessed.

From another front, privacy and security are becoming a big concern in such
networks where considerable attention from the research community has been
paid. Importantly, the broadcast nature of the wireless RF\ link renders it
vulnerable to eavesdropping attacks \cite{daniel}. While higher layers view
the security aspect as an implementation of cryptographic protocols, the
physical layer (PHY) security, introduced by Wyner, aims at establishing
secure transmissions, by ensuring that the data rate of the legitimate link
exceeds that of the wiretap one by a certain threshold.

From a multibeam satellite communication point of view, the legitimate
users-links are established through narrow multiple spot beams, where each
beam targets a single cell on the covered zone, rendering the communication
most unlikely to being intercepted from distant malicious nodes.
Nevertheless, potential eavesdroppers might be located in the same zone as
the legitimate users, and consequently, the secrecy level of the user link
is affected. To this end, several works in the literature have dealt with
the secrecy level of HTS systems as in \cite{secsat1, secsat2, secsat3},
where the analysis carried out the performance of HTS\ relay-based networks,
where the feeder link operates on RF\ spectrum. On the other hand, several
works such as \cite{lopezmartinez, oam} assessed the PHY\ security of FSO\
links.

\subsection{Motivation and Contributions}

Very few works in the literature have investigated the secrecy level of HTS\
systems with optical feeder link. In particular, the authors in \cite%
{photonics} dealt with the secrecy analysis of an HTS relay network, where
the optical link is used at the terrestrial side as a last-mile link. Also,
the wiretapper is considered only on the RF\ side. Distinctively,
contributions such as \cite{ansarisecrecy, secr4, secr5} dealt with the
secrecy level of mixed RF-FSO\ links where the RF\ link is considered in the
first hop. Furthermore, nodes parameters, a data precoding process, and an
eavesdropper on the optical link were not considered in such works. Besides,
differently from the works \cite{txdiv1,txdiv2} where the authors proposed a
transmit Laser selection diversity techniques for FSO systems, the
statistics of the combined SNR\ using receive diversity {selection combining
(SC)} scheme were not carried out in closed-form expression. Capitalizing on
this, we aim at this work to investigate the PHY\ layer security of an HTS\
multi-user relay-based system with an optical feeder link, where the
satellite, acting as a relay, converts the incoming optical wave carrying
the users' data, after combining the incoming beams to its photodetectors
through SC scheme, to an electrical signal, regenerate and conveys them to
the end-users. {Two scenarios are analyzed, namely, i) the satellite
performs ZF precoding technique, after decoding information signals on the
second hop, before transmitting it to the end-users, ii) the satellite does
not perform ZF technique and delivers the processed received signal to the
end-users}. Malicious eavesdroppers are considered on each hop of the
transmission. The main contributions of this paper can be summarized as
follows:

\begin{itemize}
\item Statistical properties of the end-to-end secrecy capacity are
retrieved, in terms of legitimate and wiretap instantaneous SNRs of the $S$-$%
R$ and $R$-$D$ hops.

\item A novel expression for the IP of dual-hop DF relaying-based systems is
retrieved, where a decoding failure event at the satellite is considered.

\item Capitalizing on the above two results, a closed-form expression of the
intercept probability (IP) of the system is derived {for the ZF\ and non-ZF\
scenarios.}

\item An asymptotic analysis of the derived analytical result is performed
based on which the achievable coding gain and diversity order are quantified.
\end{itemize}

\subsection{Organization of the Paper}

The remainder of this paper is organized as follows. Section II\ is
dedicated to present the system and channel model. Section III deals with
the statistical properties of the end-to-end SNRs, while Section IV depicts
the derived analytical expression of the IP. In Section V, illustrative
numerical results are shown to assess the effect of channel parameters on
the system's secrecy level. Finally, Section VI\ concludes the paper.

\section{System and Channel Model}

We consider in this analysis a single OGS communicating with a GEO\
satellite through an optical feeder link. The transmit OGS, acting a source (%
$S$) node, is assumed to have\ a clear line of sight (LOS) link with the
satellite unblocked by clouds. The selected source node transmits data
symbols of the $N$ users through a turbulent optical channel to the
satellite. This latter, acting as a relay ($R$), combines through $K$
optical photodetectors the incoming optical wave, converts it to the
electrical domain, performs SC technique, decodes, and delivers it in the
form of $N$\ beams to $N$\ user earth stations $\left( T_{i}\right) _{1\leq
i\leq N}$. If the received SNR\ at the satellite is greater than a
predefined decoding threshold, it can successfully decode the information
signal. Otherwise, the decoding cannot be ensured correctly. For the former
case, {t{he satellite can either forward the regenerated signals directly or
precode them using ZF technique}. We assume that an eavesdropper attempting
to overhear the divergent\ optical beam coming from the OGS to the satellite
(FSO$\ S$-$R$ hop). In addition to this, another potential wiretapper
attempts to overhear the signal carried to the earth-station $T_{i}$. }
\begin{figure}[tbp]
\centering
\hspace*{-1cm}\includegraphics[scale= .32]{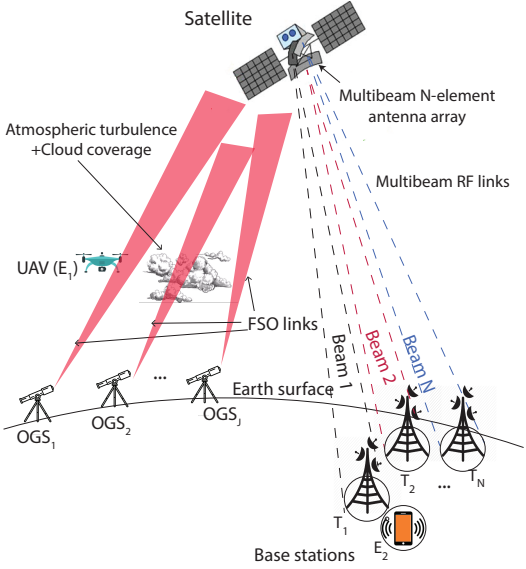} \vspace*{-1.5cm}
\caption{ {\protect\footnotesize System model.}}
\label{sysmod}
\end{figure}
\begin{figure}[tbp]
\centering
\hspace*{-1.3cm}\includegraphics[scale=.45]{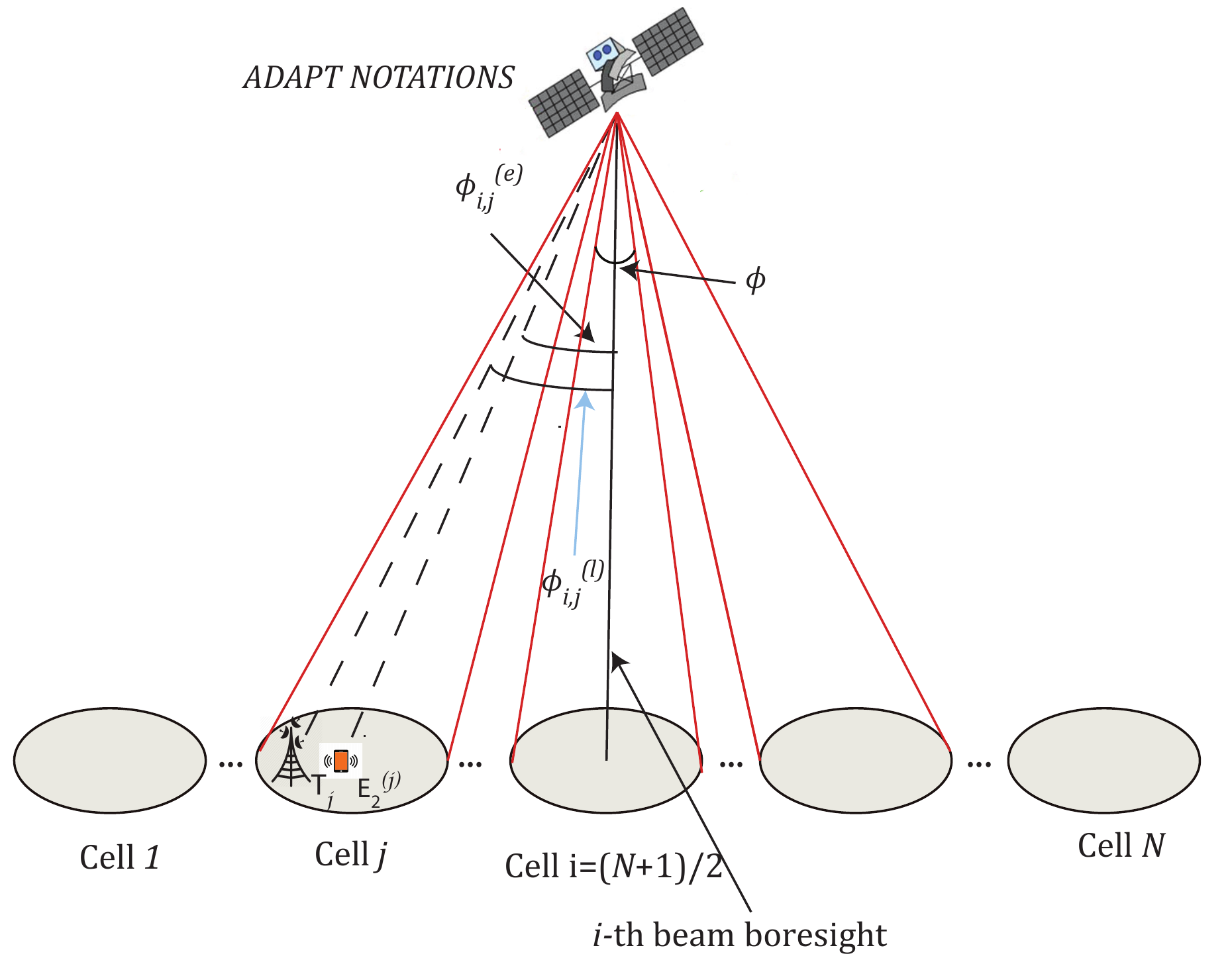}
\caption{{\protect\footnotesize Legitimate and wiretapper nodes' angles with
respect to beams' boresights.}}
\label{distr}
\end{figure}

\subsection{$S$-$R$ Link}

\subsubsection{Legitimate Link}

An OGS\ with a LOS unblocked by clouds is considered for data transmission.
Without loss of generality, all users' signals are assumed to have the same
power.

The received beam at the satellite contains $N$ multiplexed signals
transmitted from the transmit gateway. Hence, the received electrical signal
vector\ at the satellite's $k$-th aperture, after demultiplexing, is
expressed as \cite{soleimani}
\begin{equation}
\mathbf{y}_{1}^{(k)}=\sqrt{\omega _{l}P_{S}}\left( \eta I_{k}\right) ^{\frac{%
r}{2}}\mathbf{u}+\mathbf{n}_{k},k=1,\ldots ,K,
\end{equation}%
with $I_{k}=I_{k}^{(a)}I_{k}^{(p)}I_{k}^{(\ell )},$ being the product of the
irradiance fluctuation due to atmospheric turbulence, the pointing error due
to the beam misalignment, and the free-space path loss, respectively, with $%
I_{k}^{(\ell )}=I_{t}e^{-\phi d},$ and $I_{t},$ $d$, and $\phi $ denote the
laser emittance, the station-satellite distance, and the path loss exponent,
respectively. $r\in \{1,2\}$ being a detection-technique dependent
parameter, with $r=1$ referring to coherent detection, and $r=2$ stands for
direct detection, and $\omega _{l}$ denotes the portion of power received by
the satellite's photodetectors, among the total radiated power from the
selected OGS. Also:

\begin{itemize}
\item $\mathbf{u}=[u_{1},u_{2},\ldots ,u_{N}]^{T}$ denotes the transmitted
signal vector of $N$ signals, with the superscript {{$^{T}$ refering to the
transpose of a vector}}. Each unit-power signal is modulated onto one of the
$N$\ optical sub-carriers, where $\mathbb{E}\left[ \mathbf{u}^{H}\mathbf{u}%
\right] ={N},$ with $\mathbb{E}\left[ \mathbf{.}\right] $ denoting the
expectation operator.

\item $P_{S}$ is the OGS\ transmit power.

\item $\mathbf{n}_{k}=[n_{k}^{(1)},n_{k}^{(2)},\ldots ,n_{k}^{(N)}]^{T}$
stands for the additive white Gaussian noise (AWGN) process at the satellite
with zero mean and the same variance $\sigma _{1}^{2}.$
\end{itemize}

The satellite converts the received optical waves at the $K$ apertures into
electrical signals, and then uses SC to choose the branch with highest
instantaneous SNR as: $s^{\ast }=\underset{1\leq k\leq K}{\arg \max }\gamma
_{1}^{(k)},$ with $\gamma _{1}^{(k)}=\frac{P_{S}\omega _{l}\left( \eta
I_{k}\right) ^{r}}{\sigma _{1}^{2}}$ denoting the instantaneous
signal-to-noise ratio (SNR)\ received at the $k$-th satellite's aperture$.$
Consequently, the combined SNR at the satellite is: $\gamma _{1}=\underset{%
1\leq k\leq K}{\max }\gamma _{1}^{(k)}.$ Afterwards, the satellite performs
a decoding process on the combined electrical signal to regenerate the
information signal again.

\subsubsection{Wiretap Link}

The hybrid ground-satellite communication is performed under the malicious
attempt of eavesdroppers per each one of the two hops to intercept the
legitimate message. For the FSO\ link, we consider the presence of one
wiretapper $e^{(1)}$ located at some altitude within the divergence region
of the OGS beam, being able to capture a portion $\omega _{e}=1-\omega _{l}$
of the optical power. That is, the received SNR\ at $e^{(1)}$ is $\gamma
_{1}^{(e)}=\frac{\omega _{e}P_{S}\left( \eta I_{e^{(1)}}\right) ^{r}}{\sigma
_{e^{(1)}}^{2}},$ with $I_{e^{(1)}}$ and $\sigma _{e^{(1)}}^{2}$ denote the
respective optical channel gain, and the variance of the AWGN at $e^{(1)}$,
respectively.

In terrestrial FSO\ communication, the altitude-dependent refractive index
structure parameter $C_{n,\varpi }^{2}(h)$ $\left( \varpi \in \{1,e\}\right)
$ as well as the Rytov variance $\sigma _{R,\varpi }^{2}$ are two crucial
parameters that represent the atmospheric turbulence and pointing error loss
impairments. Such parameters are expressed in key system and environment
quantities. In the context of vertical optical\ links in HTS systems, the
altitude-dependent refractive structure index parameter in m$^{-\frac{2}{3}}$
and the Rytov variance can be expressed for the uplink using the
Hufnagel-Valley Boundary model as given in \cite[Eqs. (4, 9-10)]{uplink}, in
terms of the altitude $h$ in meters, the wind speed $V_{w}$ in m/s$,$ the
satellite and OGS$\ $altitudes $d_{\varpi }$ and $h_{0}$, the satellite's
zenith angle with respect to the OGS, and the operating wavelength $\lambda
. $

Interestingly, $\xi _{\varpi }^{2}=\frac{W_{eq}}{2\sigma _{s}^{2}}$ denotes
the pointing error strength, which is the ratio between the equivalent beam
waist at the satellite altitude and the beam wander displacement variance,
where the last-mentioned quantities are defined in \cite{farid}, \cite[Eq.
(2)]{rubiosat}, and \cite[Eq. (8)]{sandalidis}.

\subsection{$R$-$D$ Link}

The satellite generates $N$\ adjacent beams. The received signal vector at
the $N$ legitimate end-users\ and wiretap nodes can be formulated as%
\begin{equation}
\mathbf{y}_{2}^{(\varkappa )}=\mathbf{H}^{(\varkappa )}\mathbf{x}+\mathbf{n}%
_{\varkappa },
\end{equation}%
with $\varkappa $ equals either $l$ for the legitimate earth stations or $%
e^{(2)}$ for the second hop's wiretappers, $P_{SAT}$ is the satellite
transmit power, $\mathbf{H}^{(\varkappa )}$ is the channel matrix between
the $N$ satellite antennas and the nodes $\left[ \varkappa _{i}\right]
_{i=1,..,N}$\footnote{%
The subscripts/superscripts $"l"$ and $"e^{(2)}"$ are used to denote the
legitimate and wiretap links at the second hop, respectively.}, and $\mathbf{%
n}_{\varkappa }=\left[ n_{\varkappa }^{(i)}\right] _{i=1,..,N}$ is the AWGN\
vector whose elements are zero mean and with the same variance $\sigma
_{\varkappa }^{2}.$

It is known that the channel matrix $\mathbf{H}^{(\varkappa )}$ can be
decomposed as \cite{ahmad}%
\begin{equation}
\mathbf{H}^{(\varkappa )}=\mathbf{D}^{(\varkappa )}\mathbf{V}^{(\varkappa )},
\end{equation}%
with $\mathbf{D}^{(\varkappa )}$ being a diagonal matrix containing
real-valued random fading coefficients, and $\mathbf{V}^{(\varkappa )}$
entries are path-loss and radiation pattern coefficients, defined by $%
V_{i,j}^{(\varkappa )}=\frac{c\sqrt{G_{T}^{(j)}G_{R}^{{(\varkappa }_{i}{)}%
}a_{j}^{(\varkappa _{i})}}}{4\pi fr^{(\varkappa _{i})}\sqrt{\kappa TB_{W}}}$%
\cite{ahmad}, where $c$ being the light celerity in the free space, $%
G_{T}^{(j)}$ and $G_{R}^{(\varkappa _{i})}$ are the respective gains of the
satellite's $j$-th transmit antenna and the receive antenna of the node $%
\varkappa _{i}$, $\kappa $ is the Boltzmann constant, $T$ is the receiver
noise temperature, $f$ is the operating frequency, $r^{(\varkappa _{i})}$ is
the distance between the satellite and the node $\varkappa _{i}$, and $%
a_{j}^{(\varkappa _{i})}$ is the normalized beam radiation gain, which can
be approximated as $a_{j}^{(\varkappa _{i})}\approx \left( \frac{J_{1}\left(
\vartheta _{j}^{(\varkappa _{i})}\right) }{2\vartheta _{j}^{(\varkappa _{i})}%
}+36\frac{J_{3}\left( \vartheta _{j}^{(\varkappa _{i})}\right) }{\left(
\vartheta _{j}^{(\varkappa _{i})}\right) ^{3}}\right) ^{2}$ \cite{secsat3},
with $\vartheta _{j}^{(x_{i})}=2.07123\frac{\sin \left( \phi
_{j}^{(\varkappa _{i})}\right) }{\phi _{j,3\text{dB}}},$ and $J_{n}(.)$
denotes the $n$-th order Bessel function of the first kind \cite[Eq. (8.402)]%
{integrals}, and $\phi _{j}^{(\varkappa _{i})}$ denote the angle between the
node $\varkappa _{i}$ and the $j$-th beam boresight as indicated in (\ref%
{distr}), given as%
\begin{equation}
\phi _{j}^{(\varkappa _{i})}=\left\{
\begin{array}{c}
\phi \left( i-j\right) +\phi _{j}^{(\varkappa _{j})},\text{ }j\geq i \\
\phi \left( i-j\right) -\phi _{j}^{(\varkappa _{j})},\text{ }j<i%
\end{array}%
\right. ,  \label{phi1}
\end{equation}%
with $\phi =$ $\frac{\mathcal{D}}{r}$ referring to the beam width angle,
with $\mathcal{D}$ being the cells' diameter assumed to be equal for all
cells, and $r$ is the distance between each node $\varkappa _{i}$ and the
satellite, assumed to be equal for all nodes (i.e., $r^{(\varkappa
_{i})}\approx r,\forall i\leq N)$. Also, $\phi _{j,3\text{dB}}$ denotes the
angle corresponding to 3dB power loss of the $j$-th beam.

\subsubsection{With ZF\ Precoding}

\paragraph{Legitimate Link}

We consider the case of a ZF receiver employed in \cite{ahmad}, where the
transmit signals are precoded, after decoding, at the satellite before
transmitting them to the end-users. Thus, the transmit vector $\mathbf{x}$
is expressed as: $\mathbf{x}=\mathbf{Mu},$ where $\mathbf{M=}\sqrt{\varphi }%
\mathbf{A}$ is a $N\times N$ precoding matrix with $\mathbf{A}\mathcal{=}%
\left( \mathbf{V}^{(l)}\right) ^{-1},$ and
\begin{equation}
\varphi =\frac{P_{SAT}}{N\Tr\left[ \left( \mathbf{V}^{(l)}\left( \mathbf{V}%
^{(l)}\right) ^{H}\right) ^{-1}\right] },  \label{phiii}
\end{equation}%
where the superscript $^{H}$ refers to the Hermitian operator defined as the
conjugate of the transpose matrix, and {$\Tr[.]$} stands for the matrix's
trace.

After performing ZF\ precoding at the satellite, the received signal and
SNR\ at the $i$-th earth-station, without considering the first hop, can be
formulated as follows%
\begin{eqnarray}
y_{ZF}^{(l_{i})} &=&\sqrt{\varphi }D_{i}^{(l)}u_{i}+n^{(l_{i})}, \\
\gamma _{ZF}^{(l_{i})} &=&\overline{\gamma }_{ZF}^{(l_{i})}\left\vert
D_{i}^{(l)}\right\vert ^{2},  \label{gambar2}
\end{eqnarray}%
with $D_{i}^{(\varkappa )}$ denoting the $i$-th element of the diagonal of $%
\mathbf{D}^{(\varkappa )},$ and $\overline{\gamma }_{ZF}^{(l_{i})}=\frac{%
\varphi }{\sigma _{l}^{2}}.$ Since the satellite performs DF\ protocol, the
equivalent received SNR at the $i$-th earth-station is given as%
\begin{equation}
\gamma _{eq,ZF}^{(l_{i})}=\min \left( \gamma _{1},\gamma
_{ZF}^{(l_{i})}\right) .  \label{minimum}
\end{equation}

\paragraph{Wiretap Link}

As far as the second hop is concerned, the satellite-stations links are
ensured under the potential presence of one$\ $eavesdropper $\left(
e_{i}^{(2)}\right) _{1\leq i\leq N}$ per each cell $i$, aiming to intercept
the RF\ beam transmitted to it. The received signal and instantaneous SNR\
at $e_{i}^{(2)}$ can be expressed, respectively, as follows%
\begin{align}
y_{e_{i}^{(2)},ZF}& =\sqrt{\varphi }D_{i}^{\left( e^{(2)}\right)
}\sum_{l=1}^{N}u_{l}\sum_{j=1}^{N}V_{i,j}^{\left( e^{(2)}\right)
}A_{j,l}+n^{(e_{i}^{(2)})}, \\
\gamma _{e_{i}^{(2)},ZF}& =\frac{\psi _{i}\gamma _{ZF}^{\left(
e_{i}^{(2)}\right) }}{\theta _{i}\gamma _{ZF}^{\left( e_{i}^{(2)}\right) }+1}%
,  \label{sinr}
\end{align}%
where $n^{(e_{i}^{(2)})}$ denotes an AWGN process at $e_{i}^{(2)}$ with zero
mean and the same variance for all wiretap nodes $\sigma _{e^{(2)}}^{2},$ $%
\gamma _{ZF}^{\left( e_{i}^{(2)}\right) }=\overline{\gamma }%
_{ZF}^{(e_{i})}\left\vert D_{i}^{\left( e^{(2)}\right) }\right\vert ^{2},$ $%
\overline{\gamma }_{ZF}^{\left( e_{i}^{(2)}\right) }=$ $\frac{\varphi }{%
\sigma _{e^{(2)}}^{2}},$ $\psi _{i}=\left(
\sum\limits_{j=1}^{N}V_{i,j}^{(e^{(2)})}A_{j,i}\right) ^{2},$ and $\theta
_{i}=\left( \sum\limits_{m=1,m\neq
i}^{N}\sum\limits_{j=1}^{N}V_{i,j}^{(e^{(2)})}A_{j,m}\right) ^{2}.$

\subsubsection{Without ZF Precoding}

When no ZF\ precoding is performed at the transmit OGS, the received signal
and SNR at the legitimate earth station $l_{i}$ as well as the eavesdropper $%
e_{i}^{(2)}$ are given as \footnote{%
The subscript $"NZF"$ is used to denote the "non-ZF case", while the $"ZF"$
subscript denotes the ZF\ adoption scenario.}
\begin{eqnarray}
y_{NZF}^{(\varkappa _{i})} &=&\sqrt{\frac{P_{SAT}}{N}}D_{i}^{(\varkappa
)}\sum_{j=1}^{N}V_{i,j}^{(\varkappa )}u_{j}+n^{(\varkappa _{i})}, \\
\gamma _{\varkappa _{i},NZF} &=&\frac{\Psi ^{(\varkappa _{i})}\gamma
_{NZF}^{(\varkappa _{i})}}{\Theta ^{(\varkappa _{i})}\gamma
_{NZF}^{(\varkappa _{i})}+1},  \label{ti}
\end{eqnarray}%
with $\gamma _{NZF}^{(\varkappa _{i})}=\overline{\gamma }_{NZF}^{(\varkappa
_{i})}\left\vert D_{i}^{(\varkappa )}\right\vert ^{2},\overline{\gamma }%
_{NZF}^{(\varkappa _{i})}=\frac{P_{SAT}}{N\sigma _{\varkappa }^{2}},$ $\Psi
^{(\varkappa _{i})}=\left( V_{i,i}^{(\varkappa )}\right) ^{2},$ and $\Theta
^{(\varkappa _{i})}=\left( \sum\limits_{j=1,j\neq i}^{N}V_{i,j}^{(\varkappa
)}\right) ^{2}$.

\section{Statistical properties}

In this section, the\ cumulative distribution function (CDF)\ of the SNR\ of
the legitimate link both hops, as well as the wiretap link, is expressed in
terms of the system and channel parameters.

\subsection{$S$-$R$ Link}

The Gamma-Gamma distribution is considered for modeling the atmospheric
turbulence induced-fading with pointing error for each received optical
beam. The respective probability density function (PDF)\ and CDF\ of the
SNRs $\gamma _{1}^{(k)}$ and $\gamma _{1}^{(e)},$ received respectively at
the satellite's $k$-th aperture and the received one at $E_{1},$ are given
as \cite{zedini}
\begin{align}
f_{\gamma _{1}^{(\varpi )}}(z)& =\frac{\mathcal{P}_{\varpi }}{rz}%
G_{1,3}^{3,0}\left( \Upsilon _{\varpi }\left( \frac{z}{\mu _{r}^{(\varpi )}}%
\right) ^{\frac{1}{r}}\left\vert
\begin{array}{c}
-;\xi _{\varpi }^{2}+1 \\
\xi _{\varpi }^{2},\alpha _{\varpi },\beta _{\varpi };-%
\end{array}%
\right. \right) ,  \label{pdfx} \\
F_{\gamma _{1}^{(\varpi )}}(z)& =\frac{\left( 2\pi \right) ^{1-r}\mathcal{P}%
_{\varpi }}{r^{2-\alpha _{\varpi }-\beta _{\varpi }}}G_{r+1,3r+1}^{3r,1}%
\left( \frac{\Upsilon _{\varpi }^{r}z}{r^{2r}\mu _{r}^{(\varpi )}}\left\vert
\begin{array}{c}
1;\kappa _{1}^{(\varpi )} \\
\kappa _{2}^{(\varpi )};0%
\end{array}%
\right. \right) ,  \label{cdfx}
\end{align}%
respectively, with $\Upsilon _{\varpi }=\frac{\xi _{\varpi }^{2}\alpha
_{\varpi }\beta _{\varpi }}{\xi _{\varpi }^{2}+1},\varpi \in \{k,e\},$ $%
\mathcal{P}_{\varpi }=\frac{\xi _{\varpi }^{2}}{\Gamma \left( \alpha
_{\varpi }\right) \Gamma \left( \beta _{\varpi }\right) },$ $\mu
_{r}^{(\varpi )}=\mathbb{E}\left[ \gamma _{1}^{(\varpi )}\right] $, $\kappa
_{1}^{(\varpi )}=\left( \frac{\xi _{\varpi }^{2}+i}{r}\right) _{i=1,..,r},$ $%
\kappa _{2}^{(\varpi )}=\left( \frac{\xi _{\varpi }^{2}+i}{r},\frac{\alpha
_{\varpi }+i}{r},\frac{\beta _{\varpi }+i}{r}\right) _{i=0,..,r-1}$, and $%
G_{p,q}^{m,n}\left( .\left\vert .\right. \right) $ refers to the Meijer's $G$%
-function \cite[Eqs. (1.111), (1.112)]{mathai}. Moreover, the
turbulence-induced fading parameters $\alpha _{\varpi }$ and $\beta _{%
\mathcal{\ \varpi }}$ are expressed terms of the Rytov variance $\sigma
_{R,\varpi }^{2}$ given in \cite[Eqs. (4, 9-10)]{uplink} using \cite[Eqs.
(9-10)]{ma}. Without loss of generality, we consider that the fading
amplitudes are i.i.d on all branches, that is $\alpha _{k}=\alpha _{1}$, $%
\beta _{k}=\beta _{1}$, $\mu _{1}^{(k)}=\mu _{1},$ $\xi _{k}=\xi _{1}$ for $%
k=1,..,K.$

\begin{proposition}
The CDF\ of the combined SNR\ at the satellite can be expressed as follows%
\begin{equation}
F_{\gamma _{1}}(z)=\mathcal{P}_{1}^{K}\sum_{h_{1}+h_{2}+h_{3}=K}\sum%
\limits_{l=0}^{\infty }\frac{\mathcal{F}_{h_{1},h_{2},h_{3},l}}{\left(
\Upsilon z\right) ^{-\varrho _{l,h_{1},h_{2},h_{3}}}},  \label{cdfgam1}
\end{equation}%
with $\mathcal{F}_{h_{1},h_{2},h_{3},l}=\frac{K!}{h_{1}!h_{2}!h_{3}!}\left(
a_{0}^{(1)}\right)
^{h_{1}}\sum\limits_{q_{2}+q_{3}=l}c_{q_{2}}^{(2)}c_{q_{3}}^{(3)},$ $%
a_{0}^{(1)}=\frac{\Gamma \left( \alpha _{1}-\xi _{1}^{2}\right) \Gamma
\left( \beta _{1}-\xi _{1}^{2}\right) }{\xi _{1}^{2}}$, $\Upsilon =\frac{\xi
_{1}^{2}\alpha _{1}\beta _{1}}{\mu _{1}\left( \xi _{1}^{2}+1\right) },$ $%
\varrho _{l,h_{1},h_{2},h_{3}}=l+h_{1}\xi _{1}^{2}+h_{2}\alpha
_{1}+h_{3}\beta _{1}$,
\begin{equation}
c_{m}^{(i)}=\left\{
\begin{array}{l}
\left( a_{0}^{(i)}\right) ^{h_{i}},m=0 \\
\frac{1}{ma_{0}}\sum\limits_{j=1}^{m}\left( jh_{i}-m+j\right)
a_{j}^{(i)}c_{m-j}^{(i)};m\geq 1%
\end{array}%
\right. ,
\end{equation}%
and $a_{l}^{(2)}=\frac{(-1)^{l}\Gamma \left( \beta _{1}-\alpha _{1}-l\right)
}{l!\left( \xi _{1}^{2}-\alpha _{1}-l\right) \left( l+\alpha _{1}\right) }%
,a_{l}^{(3)}=\frac{(-1)^{l}\Gamma \left( \alpha _{1}-\beta _{1}-l\right) }{%
l!\left( \xi _{1}^{2}-\beta _{1}-l\right) \left( l+\beta _{1}\right) }.$
\begin{IEEEproof}
The proof is provided in Appendix A.
\end{IEEEproof}
\end{proposition}

\subsection{$R$-$D$ Link}

\subsubsection{With ZF Precoding}

On the other hand, shadowed-Rician fading channel is considered for the
satellite RF\ links, where the PDF\ of the SNR\ $\gamma _{ZF}^{(\varkappa
_{i})}$ can be calculated by applying Jacobi transform on the fading
envelope PDF given in \cite{abdi} as follows
\begin{equation}
f_{\gamma _{ZF}^{(\varkappa _{i})}}\left( z\right) =\frac{\lambda
_{\varkappa _{i}}}{\overline{\gamma }_{ZF}^{(\varkappa _{i})}}e^{-\frac{\rho
_{\varkappa _{i}}z}{\overline{\gamma }_{ZF}^{(\varkappa _{i})}}}\text{ }%
_{1}F_{1}\left( m_{s}^{(\varkappa _{i})};1;\frac{\delta _{\varkappa _{i}}z}{%
\overline{\gamma }_{ZF}^{(\varkappa _{i})}}\right) ,  \label{pdfgam2}
\end{equation}%
with$\ _{1}F_{1}(.;.;.)$ denoting the confluent hypergeometric function \cite%
[Eq. (9.210)]{integrals}, and \newline
$\lambda _{\varkappa _{i}}=\frac{1}{2b_{\varkappa _{i}}}\left( \frac{%
2b_{\varkappa _{i}}m_{s}^{(\varkappa _{i})}}{2b_{\varkappa
_{i}}m_{s}^{(\varkappa _{i})}+\Omega _{s}^{(\varkappa _{i})}}\right)
^{m_{s}^{(\varkappa _{i})}},$ $\rho _{\varkappa _{i}}=\frac{1}{2b_{\varkappa
_{i}}},$ $\delta _{\varkappa _{i}}=\frac{\Omega _{s}^{(\varkappa _{i})}}{%
2b_{\varkappa _{i}}\left( 2bm_{s}^{(\varkappa _{i})}+\Omega _{s}^{(\varkappa
_{i})}\right) }.$ Also, $\Omega _{s}^{(\varkappa _{i})}$, $2b_{\varkappa
_{i}},$ and $m_{s}^{(\varkappa _{i})}$ stand for the average power of LOS
and multipath components, and the fading severity parameter, respectively.

Consequently, by {{representing the hypergeometric function $_{1}F_{1}\left(
.;.;.\right) $ through the finite series \cite[Eq. (9)]{bankey} for integer
parameter values, and}} using {{\cite[Eq. (3.351.1)]{integrals}}}, the
respective CDF is expressed as
\begin{equation}
F_{\gamma _{ZF}^{(\varkappa _{i})}}\left( z\right) =\lambda _{\varkappa
_{i}}\sum\limits_{n=0}^{m_{s}^{(\varkappa _{i})}-1}\binom{m_{s}^{(\varkappa
_{i})}-1}{n}\frac{\delta _{\varkappa _{i}}^{n}}{v_{\varkappa _{i}}^{n+1}n!}%
\gamma _{inc}\left( n+1,\frac{v_{\varkappa _{i}}z}{\overline{\gamma }%
_{ZF}^{(\varkappa _{i})}}\right) ,  \label{cdfgam2}
\end{equation}%
{{where $\gamma _{inc}\left( .,.\right) $ stands for the lower incomplete
Gamma function \cite[Eq. (8.350.1)]{integrals}, with $v_{\varkappa
_{i}}=\rho _{\varkappa _{i}}-\delta _{\varkappa _{i}}$.}} Furthermore, the
CDF and PDF\ of the end-to-end SINR in (\ref{sinr}) at $e_{i}^{(2)}$ are
expressed as
\begin{align}
F_{\gamma _{e_{i}^{(2)},ZF}}\left( z\right) & =\left\{
\begin{array}{l}
F_{\gamma _{ZF}^{\left( e_{i}^{(2)}\right) }}\left( \frac{z}{\psi
_{i}-\theta _{i}z}\right) ;z<\mathscr{L}_{i} \\
1;z\geq \mathscr{L}_{i}%
\end{array}%
\right. ,  \notag \\
f_{\gamma _{e_{i}^{(2)},ZF}}\left( z\right) & =\left\{
\begin{array}{l}
\frac{\psi _{i}}{\left( \psi _{i}-\theta _{i}z\right) ^{2}}\frac{\partial
F_{\gamma _{ZF}^{\left( e_{i}^{(2)}\right) }}\left( \frac{z}{\psi
_{i}-\theta _{i}z}\right) }{\partial z};z<\mathscr{L}_{i} \\
0;z\geq \mathscr{L}_{i}%
\end{array}%
\right. .  \label{cdfeve2}
\end{align}%
with $\mathscr{L}_{i}=\frac{\psi _{i}}{\theta _{i}}$.

\subsubsection{Without ZF\ Precoding}

Importantly, when no ZF precoding is performed, the CDF\ of the received
SNR\ at node $\varkappa _{i},$ given in (\ref{ti}), can be expressed as
\begin{equation}
F_{\gamma _{\varkappa _{i},NZF}}\left( z\right) =\Pr \left[ \gamma
_{NZF}^{(\varkappa _{i})}\left( \Psi ^{(\varkappa _{i})}-\Theta ^{(\varkappa
_{i})}z\right) <z\right] ,
\end{equation}%
where two cases are distinguished, namely $z<\mathcal{L}^{(\varkappa _{i})}$
and $z\geq \mathcal{L}^{(\varkappa _{i})},$ with $\mathcal{L}^{(\varkappa
_{i})}=\frac{\Psi ^{(\varkappa _{i})}}{\Theta ^{(\varkappa _{i})}}.$
Similarly to the ZF precoding\ case, the CDF and PDF\ of the received SNR\
at node $\varkappa _{i}$ are given as {\small
\begin{align}
F_{\gamma _{\varkappa _{i},NZF}}\left( z\right) & =\left\{
\begin{array}{l}
F_{\gamma _{NZF}^{(\varkappa _{i})}}\left( \frac{z}{\Psi ^{(\varkappa
_{i})}-\Theta ^{(\varkappa _{i})}z}\right) ;z<\mathcal{L}^{(\varkappa _{i})}
\\
1;z\geq \mathcal{L}^{(\varkappa _{i})}%
\end{array}%
\right. ,  \notag \\
\text{ }f_{\gamma _{\varkappa _{i},NZF}}\left( z\right) & =\left\{
\begin{array}{l}
\frac{\Psi ^{(\varkappa _{i})}}{\left( \Psi ^{(\varkappa _{i})}-\Theta
^{(\varkappa _{i})}z\right) ^{2}}\frac{\partial F_{\gamma _{NZF}^{(\varkappa
_{i})}}\left( \frac{z}{\Psi ^{(\varkappa _{i})}-\Theta ^{(\varkappa _{i})}z}%
\right) }{\partial z};z<\mathcal{L}^{(\varkappa _{i})} \\
0;z\geq \mathcal{L}^{(\varkappa _{i})}%
\end{array}%
\right. .  \label{cdfnzf}
\end{align}%
} with $F_{\gamma _{NZF}^{(\varkappa _{i})}}\left( .\right) $ is obtained
from (\ref{cdfgam2}) by replacing $\overline{\gamma }_{ZF}^{(\varkappa
_{i})} $ with $\overline{\gamma }_{NZF}^{(\varkappa _{i})}$.

\section{Secrecy performance analysis}

The intercept probability metric is defined as the probability that the
secrecy capacity, which is the difference between the capacity of the
legitimate links and that of the wiretap channels (i.e., $\{S$-$R,$ $S$-$%
e^{(1)}\},$ $\{R$-$l_{i},$ $R$-$e_{i}^{(2)}\})$, equals to zero.
Additionally, from \cite[Eq. (26)]{lapidoth} and \cite{chaaban}, the
respective channel capacities of the FSO legitimate and wiretap links are
given for coherent detection techniques as%
\begin{equation}
C_{\varrho }=\log _{2}\left( 1+Y_{\varrho }\right) ,
\end{equation}%
with $\left( \varrho ,Y_{\varrho }\right) $ equals either $\left( 1,\gamma
_{1}\right) $ or $\left( e^{(1)},\gamma _{1}^{(e)}\right) .$

On the other hand, the system's intercept probability is defined as%
\begin{align}
P_{int,\Xi }^{(i)}& =\Pr \left( C_{s,\Xi }^{(i)}=0\right)  \notag \\
& =\Pr \left( \left. C_{s,\Xi }^{(i)}=0\right\vert \gamma _{1}>\gamma
_{th}\right) \Pr \left( \gamma _{1}>\gamma _{th}\right)  \notag \\
& +\Pr \left( \left. C_{s,\Xi }^{(i)}=0\right\vert \gamma _{1}<\gamma
_{th}\right) \Pr \left( \gamma _{1}<\gamma _{th}\right) ,  \label{ipnew}
\end{align}%
where $\gamma _{th}$ is a decoding threshold SNR, below which the decoding
process fails, {$\Xi $ }${\in \{}${\ $ZF,NZF\},$ and}%
\begin{equation}
C_{s,\Xi }^{(i)}=\min \left( C_{s}^{(1)},C_{s,eq,\Xi }^{(2,i)}\right) ,
\label{cs1}
\end{equation}%
with%
\begin{equation}
C_{s}^{(1)}=\log _{2}\left( \frac{1+\gamma _{1}}{1+\gamma _{1}^{(e)}}\right)
,  \label{cs11}
\end{equation}%
and%
\begin{equation}
C_{s,eq,\Xi }^{(2,i)}=\min \left( C_{s,\Xi }^{(1,2,i)},C_{s,\Xi
}^{(2,i)}\right) ,  \label{cseq}
\end{equation}%
stand for the end-to-end, first hop, and second hops' secrecy capacities,
respectively, considering DF\ relaying protocol, with%
\begin{eqnarray}
C_{s,\Xi }^{(1,2,i)} &=&\log _{2}\left( \frac{1+\gamma _{1}}{1+\gamma
_{e_{i}^{(2)},\Xi }}\right) ,  \label{cs12} \\
C_{s,\Xi }^{(2,i)} &=&\log _{2}\left( \frac{1+\gamma _{l_{i},\Xi }}{1+\gamma
_{e_{i}^{(2)},\Xi }}\right) .  \label{cs2}
\end{eqnarray}

{Therefore, the overall secrecy capacity in (\ref{cs1}) can be expressed as}%
\begin{equation}
C_{s,\Xi }^{(i)}=\min \left( C_{s}^{(1)},C_{s,\Xi }^{(2,i)},C_{s,\Xi
}^{(1,2,i)}\right) .  \label{csmin}
\end{equation}

\begin{lemma}
The system's intercept probability for DF\ relaying scheme is given as%
\begin{equation}
P_{int,\Xi }^{(i)}=1-\int_{y=\gamma _{th}}^{\infty }f_{\gamma _{1}}\left(
y\right) F_{\gamma _{1}^{(e)}}\left( y\right) \mathcal{J}_{\Xi }(y)dy,
\label{propip}
\end{equation}%
where $\mathcal{J}_{\Xi }(y)=\int_{z=0}^{y}f_{\gamma _{e_{i}^{(2)},\Xi
}}\left( z\right) F_{\gamma _{l_{i},\Xi }}^{c}\left( z\right) dz,$ with $%
F_{.}^{c}\left( .\right) $ accounts for the complementary CDF \footnote{%
Both notations $\gamma _{l_{i},ZF}$ and $\gamma _{ZF}^{\left( l_{i}\right) }$
stand for the same random variable.}.
\begin{IEEEproof}
The proof is provided in Appendix B.
\end{IEEEproof}
\end{lemma}

\subsection{Exact Analysis}

\begin{lemma}
The integral $\mathcal{J}_{\Xi }(y)$ in the expression {(\ref{propip}) is
given for the ZF and NZF cases as }%
\begin{align}
\mathcal{J}_{ZF}(y)&
=\sum\limits_{n_{1}=0}^{m_{s}^{(l_{i})}-1}\sum\limits_{n_{2}=0}^{m_{s}^{%
\left( e_{i}^{(2)}\right) }-1}\frac{\mathcal{U}_{i}\left( n_{1},n_{2}\right)
}{\exp \left( r_{1}^{\left( e_{i}^{(2)}\right) }\right) }\sum%
\limits_{k_{1}=0}^{n_{1}}\sum\limits_{j=0}^{\infty }\frac{(-1)^{j}}{j!k_{1}!}
\notag \\
& \times \left( \psi _{i}r_{1}^{(l_{i})}r_{1}^{\left( e_{i}^{(2)}\right)
}\right) ^{j+k_{1}}\sum\limits_{p=0}^{k_{1}+n_{2}+j}\frac{\binom{%
k_{1}+n_{2}+j}{p}}{\left( -r_{1}^{\left( e_{i}^{(2)}\right) }\right)
^{p-n_{2}}}  \notag \\
& \times \left[
\begin{array}{c}
\theta _{i}r_{1}^{\left( e_{i}^{(2)}\right) }\Gamma \left(
p+1-k_{1}-j,r_{1}^{\left( e_{i}^{(2)}\right) }\right)  \\
-\Gamma \left( p+1-k_{1}-j,\frac{r_{1}^{\left( e_{i}^{(2)}\right) }\psi _{i}%
}{\psi _{i}-\theta _{i}y}\right)
\end{array}%
\right] ,  \label{lemma21}
\end{align}%
\begin{figure*}[t]
{\normalsize 
\setcounter{mytempeqncnt}{\value{equation}}
\setcounter{equation}{31} }
\par
\begin{equation}
\mathcal{J}_{NZF}(y)=\left\{
\begin{array}{l}
\sum\limits_{n_{1}=0}^{m_{s}^{(l_{i})}-1}\sum\limits_{n_{2}=0}^{m_{s}^{%
\left( e_{i}^{(2)}\right) }-1}\frac{\mathcal{U}_{i}\left( n_{1},n_{2}\right)
}{\exp \left( -\mathcal{G}^{\left( l_{i},e_{i}^{(2)}\right) }\right) }%
\sum\limits_{j=0}^{\infty }\frac{\left( -1\right) ^{j}\left( \mathcal{G}%
^{\left( e_{i}^{(2)},l_{i}\right) }\right) ^{j+n_{2}+1}}{k_{1}!j!}%
\sum\limits_{p=0}^{n_{2}+k_{1}+j}\frac{\binom{n_{2}+k_{1}+j}{p}}{\left( -%
\mathcal{G}^{\left( l_{i},e_{i}^{(2)}\right) }\right) ^{p-k_{1}-j-n_{2}-1}}
\\
\times \left[
\begin{array}{l}
\Gamma \left( p-n_{2}-1-j,\mathcal{G}^{\left( l_{i},e_{i}^{(2)}\right)
}\right) \\
-\Gamma \left( p-n_{2}-1-j,\frac{\mathcal{G}^{(l_{i},l_{i})}\mathcal{T}_{i}}{%
\Theta ^{(l_{i})}\left( \Psi ^{(l_{i})}-\Theta ^{(l_{i})}y\right) }+\frac{%
\mathcal{G}^{\left( l_{i},e_{i}^{(2)}\right) }\Theta ^{\left(
e_{i}^{(2)}\right) }}{\Theta ^{(l_{i})}}\right)%
\end{array}%
\right] ,\mathcal{L}^{(l_{i})}<\mathcal{L}^{\left( e_{i}^{(2)}\right) } \\
\sum\limits_{n_{1}=0}^{m_{s}^{(l_{i})}-1}\sum\limits_{n_{2}=0}^{m_{s}^{%
\left( e_{i}^{(2)}\right) }-1}\frac{\mathcal{U}_{i}\left( n_{1},n_{2}\right)
}{\exp \left( \mathcal{G}^{\left( e_{i}^{(2)},l_{i}\right) }\right) }%
\sum\limits_{k_{1}=0}^{n_{1}}\sum\limits_{j=0}^{\infty }\frac{%
(-1)^{k_{1}}\left( \mathcal{G}^{\left( l_{i},e_{i}^{(2)}\right) }\right)
^{j+k_{1}}}{k_{1}!j!}\sum\limits_{p=0}^{n_{2}+k_{1}+j}\frac{\binom{%
n_{2}+k_{1}+j}{p}}{\left( \mathcal{G}^{\left( e_{i}^{(2)},l_{i}\right)
}\right) ^{p-n_{2}-j-k_{1}}} \\
\times \left[
\begin{array}{l}
\Gamma \left( p-k_{1}-j+1,-\mathcal{G}^{\left( e_{i}^{(2)},l_{i}\right)
}\right) \\
-\Gamma \left( p-k_{1}-j+1,\left( \frac{\mathcal{T}_{i}}{\Psi ^{\left(
e_{i}^{(2)}\right) }-\Theta ^{\left( e_{i}^{(2)}\right) }y}-\Theta
^{(l_{i})}\right) \frac{\mathcal{G}^{\left( e_{i}^{(2)},e_{i}^{(2)}\right) }%
}{\Theta ^{\left( e_{i}^{(2)}\right) }}\right)%
\end{array}%
\right] ,\mathcal{L}^{(l_{i})}>\mathcal{L}^{\left( e_{i}^{(2)}\right) }%
\end{array}%
\right.  \label{lemma22}
\end{equation}%
\par
{\normalsize 
\hrulefill 
\vspace*{4pt} }
\end{figure*}
and in ({\ref{lemma22}}) at the top of the next page {\ for }$y<s_{i}$, with
$s_{i}$ equals $\mathscr{L}_{i}$ or $\mathcal{L}^{\left( e_{i}^{(2)}\right) }
$ for the ZF\ and NZF\ cases, respectively, $r_{1}^{(\varkappa _{i})}=\frac{%
v_{\varkappa _{i}}}{\overline{\gamma }_{2,ZF}^{(\varkappa _{i})}\theta _{i}},
$ $\mathcal{G}^{(a_{i},b_{i})}=\frac{v_{a_{i}}\Psi _{i}^{(b)}}{\mathcal{T}%
_{i}\overline{\gamma }_{2,NZF}^{(a_{i})}},\left( a,b\in \{l,e^{(2)}\}\right)
,$ $\mathcal{T}_{i}=\Psi ^{\left( e_{i}^{(2)}\right) }\Theta ^{(l_{i})}-\Psi
^{(l_{i})}\Theta ^{\left( e_{i}^{(2)}\right) },$ $\Gamma \left( .,.\right) $
is the upper-incomplete Gamma function \cite[Eq. (8.350.2)]{integrals}, and

$\mathcal{U}_{i}\left( n_{1},n_{2}\right) \mathcal{=}\frac{\lambda
_{l_{i}}\lambda _{e_{i}^{(2)}}}{n_{2}!}\left(
\begin{array}{c}
m_{s}^{(l_{i})}-1 \\
n_{1}%
\end{array}%
\right) \left(
\begin{array}{c}
m_{s}^{\left( e_{i}^{(2)}\right) }-1 \\
n_{2}%
\end{array}%
\right) \frac{\delta _{l_{i}}^{n_{1}}}{v_{l_{i}}^{n_{1}+1}}\frac{\delta
_{e_{i}^{(2)}}^{n_{2}}}{v_{e_{i}^{(2)}}^{n_{2}+1}}.$
\end{lemma}

\begin{remark}
It is noteworthy that the system's IP will be computed for the ZF\ precoding
scenario for two cases, namely $\mathscr{L}_{i}<\gamma _{th}$ and $%
\mathscr{L}_{i}\geq \gamma _{th},$ while for the non-ZF\ scenario, the IP\
expression is limited only to the case when $\mathcal{L}^{\left(
e_{i}^{(2)}\right) }<\gamma _{th},$ due to the toughness of the encountered
mathematical computation.%
\begin{IEEEproof}
The proof is provided in Appendix C.
\end{IEEEproof}
\end{remark}

\begin{lemma}
The integral $K\left( \varphi \right) =\int_{y=\varphi }^{\infty }f_{\gamma
_{1}}\left( y\right) F_{\gamma _{1}^{(e)}}dy$ can be expressed as follows
\begin{equation}
K\left( \varphi \right) =1-F_{\gamma _{1}}\left( \varphi \right) F_{\gamma
_{1}^{(e)}}\left( \varphi \right) -\mathcal{H}_{1}+\mathcal{H}_{2}\left(
\varphi \right) ,  \label{kphi}
\end{equation}%
with
\begin{equation}
\mathcal{H}_{1}=\frac{\xi _{e}^{2}}{\mathcal{P}_{1}^{-K}}%
\sum_{h_{1}+h_{2}+h_{3}=K}\sum\limits_{l=0}^{\infty }\mathcal{F}%
_{h_{1},h_{2},h_{3},l}\frac{\left( \alpha _{e}\right) _{\varrho
_{l,h_{1},h_{2},h_{3}}}\left( \beta _{e}\right) _{\varrho
_{l,h_{1},h_{2},h_{3}}}}{\left( \xi _{e}^{2}+\varrho
_{l,h_{1},h_{2},h_{3}}\right) \Upsilon _{e}^{\varrho _{l,h_{1},h_{2},h_{3}}}}%
,  \label{H1}
\end{equation}%
\begin{align}
\mathcal{H}_{2}\left( \varphi \right) & =\mathcal{P}_{1}^{K}\mathcal{P}%
_{e}\sum_{h_{1}+h_{2}+h_{3}=K}\sum\limits_{l=0}^{\infty }\frac{\varrho
_{l,h_{1},h_{2},h_{3}}\mathcal{F}_{h_{1},h_{2},h_{3},l}}{\varphi ^{-\varrho
_{l,h_{1},h_{2},h_{3}}}}  \notag \\
& \times G_{3,5}^{3,2}\left( \Upsilon _{e}\varphi \left\vert
\begin{array}{c}
1-\varrho _{l,h_{1},h_{2},h_{3}},1;\xi _{e}^{2}+1 \\
\xi _{e}^{2},\alpha _{e},\beta _{e};0,-\varrho _{l,h_{1},h_{2},h_{3}}%
\end{array}%
\right. \right) ,  \label{H2}
\end{align}%
where $\left( .\right) _{.}$ stands for the Pochhammer symbol \cite[Eq.
(06.10.02.0001.01)]{wolfram}, and $\Upsilon _{e}=\frac{\xi _{e}^{2}\alpha
_{e}\beta _{e}}{\mu _{1}^{(e)}\left( \xi _{e}^{2}+1\right) }.$%
\begin{IEEEproof}
The proof is provided in Appendix D.
\end{IEEEproof}
\end{lemma}

\begin{proposition}
The system's IP closed-form expression is given in (\ref{ipclosedform}) at
the top of the next page, as well as
\begin{figure*}[t]
{\normalsize 
\setcounter{mytempeqncnt}{\value{equation}}
\setcounter{equation}{35} }
\par
\begin{equation}
P_{int,ZF}^{(i)}=\left\{
\begin{array}{l}
1-\mathcal{J}_{ZF}\left( \mathscr{L}_{i}\right) \mathcal{K}\left( \gamma
_{th}\right) ;\mathscr{L}_{i}<\gamma _{th} \\
1-\mathcal{O}_{1}^{(i)}-\mathcal{O}_{2}^{(i)}-\mathcal{J}_{ZF}\left( %
\mathscr{L}_{i}\right) \mathcal{K}\left( \mathscr{L}_{i}\right) ;\gamma
_{th}\leq \mathscr{L}_{i}<2\gamma _{th} \\
1-\mathcal{O}_{1}^{(i)}-\mathcal{O}_{3}^{(i)}-\mathcal{J}_{ZF}\left( %
\mathscr{L}_{i}\right) \mathcal{K}\left( \mathscr{L}_{i}\right) ;\mathscr{L}%
_{i}>2\gamma _{th}%
\end{array}%
\right. ,  \label{ipclosedform}
\end{equation}%
\par
{\normalsize 
\hrulefill 
\vspace*{4pt} }
\end{figure*}
\begin{equation}
P_{int,NZF}^{(i)}=1-\mathcal{J}_{NZF}\left( \min \left[ \mathcal{L}%
^{(l_{i})},\mathcal{L}^{\left( e_{i}^{(2)}\right) }\right] \right) \mathcal{K%
}\left( \gamma _{th}\right) ,  \label{ipclosedform2}
\end{equation}%
for $\mathcal{L}^{\left( e_{i}^{(2)}\right) }<\gamma _{th},$ with%
\begin{equation}
\mathcal{O}_{1}^{(i)}=\mathcal{J}_{ZF}\left( \mathscr{L}_{i}\right) \left[
\mathcal{K}\left( \gamma _{th}\right) -\mathcal{K}\left( \mathscr{L}%
_{i}\right) \right] ,  \label{O1}
\end{equation}%
\begin{align}
\mathcal{O}_{m+1}^{(i)}& =\frac{\mathcal{P}_{1}^{K}\mathcal{P}_{e}}{\exp
\left( -r_{1}^{\left( e_{i}^{(2)}\right) }\right) }%
\sum_{h_{1}+h_{2}+h_{3}=K}\sum\limits_{l=0}^{\infty }\mathcal{F}%
_{h_{1},h_{2},h_{3},l}\varrho _{l,h_{1},h_{2},h_{3}}  \notag \\
& \times
\sum\limits_{n_{1}=0}^{m_{s}^{(l_{i})}-1}\sum\limits_{n_{2}=0}^{m_{s}^{%
\left( e_{i}^{(2)}\right) }-1}\mathcal{U}_{i}\left( n_{1},n_{2}\right)
\sum_{k_{1}=0}^{n_{1}}\sum\limits_{j=0}^{\infty }\frac{\left(
r_{1}^{(l_{i})}\psi _{i}\right) ^{j+k_{1}}}{j!k_{1}!}  \notag \\
& \times \sum\limits_{p=0}^{k_{1}+n_{2}+j}\frac{\binom{k_{1}+n_{2}+j}{p}%
\left[ \mathcal{B}_{1}^{(m,i)}\mathcal{+B}_{\alpha _{e}}^{(m,i)}\mathcal{+B}%
_{\beta _{e}}^{(m,i)}\right] }{\left( -r_{1}^{\left( e_{i}^{(2)}\right)
}\right) ^{p-n_{2}-k_{1}-j}};m=1,2,  \label{Om}
\end{align}%
and%
\begin{eqnarray}
\mathcal{B}_{\xi _{e}^{2}}^{(m,i)} &=&\left\{
\begin{array}{l}
\mathcal{S}_{i}\left( \xi _{e}^{2},0\right) ;m=1 \\
\mathcal{R}_{i}\left( \xi _{e}^{2},0\right) ;m=2%
\end{array}%
\right. ,  \notag \\
\mathcal{B}_{x}^{(m,i)} &=&\left\{
\begin{array}{l}
\sum\limits_{v=0}^{\infty }\mathcal{S}_{i}\left( x,v\right) ;m=1 \\
\sum\limits_{v=0}^{\infty }\mathcal{R}_{i}\left( x,v\right) ;m=2%
\end{array}%
\right. ,x\in \left\{ \alpha _{e},\beta _{e}\right\} ,  \label{B1final}
\end{eqnarray}%
with $b_{\xi _{e}^{2}}^{(0)}=\frac{\Gamma \left( \alpha _{e}-\xi
_{e}^{2}\right) \Gamma \left( \beta _{e}-\xi _{e}^{2}\right) }{\xi _{e}^{2}},
$ $b_{\alpha _{e}}^{(v)}=\frac{(-1)^{v}\Gamma \left( \beta _{e}-\alpha
_{e}-v\right) }{v!\left( \xi _{e}^{2}-\alpha _{e}-v\right) \left( v+\alpha
_{e}\right) },b_{\beta _{e}}^{(v)}=\frac{(-1)^{v}\Gamma \left( \alpha
_{e}-\beta _{e}-v\right) }{v!\left( \xi _{e}^{2}-\beta _{e}-v\right) \left(
l+\beta _{e}\right) }$,

$\mathcal{C}(x,q,v)=(-1)^{q}\left( g(x,q,v)-1\right) q!,$ $%
g(x,q,v)=1+q+v+x+\varrho _{l,h_{1},h_{2},h_{3}},$%
\begin{align}
\mathcal{S}_{i}\left( y,v\right) & =\frac{\Upsilon _{e}^{y+v}}{\mathscr{L}%
_{i}^{-\mathcal{C}(y,0,v)}}b_{x}^{(v)}\sum\limits_{q=0}^{\infty }\frac{%
\left( 1-\frac{\mathscr{L}_{i}}{\gamma _{th}}\right) ^{q+1}}{q!}  \notag \\
& \times G_{0,3}^{4,0}\left( \left. \frac{r_{1}^{\left( e_{i}^{(2)}\right)
}\gamma _{th}}{\mathscr{L}_{i}-\gamma _{th}}\right\vert \Lambda _{1}\left(
y,q,v\right) \right) ,  \label{si}
\end{align}%
\begin{align}
\mathcal{R}_{i}\left( y,v\right) & =\frac{\Upsilon _{e}^{y+v}b_{y}^{(v)}}{%
\mathscr{L}_{i}^{-\mathcal{C}(y,0,v)}}\sum\limits_{q=0}^{\infty }\frac{%
\left( \frac{\mathscr{L}_{i}}{\gamma _{th}}-1\right) ^{-\mathcal{C}(y,0,v)-q}%
}{\mathcal{C}(y,q,v)}  \notag \\
& \times G_{0,2}^{3,0}\left( \left. r_{1}^{\left( e_{i}^{(2)}\right)
}\right\vert \Lambda _{2}\left( y,q,v\right) \right) -\frac{\Upsilon
_{e}^{y+v}b_{y}^{(v)}\Gamma \left( \mathcal{C}(y,0,v)\right) }{\mathscr{L}%
_{i}^{-\mathcal{C}(\alpha _{e},0,v)}}  \notag \\
& \times G_{0,1}^{2,0}\left( r_{1}^{\left( e_{i}^{(2)}\right) }\left\vert
\begin{array}{c}
-;g(y,0,v) \\
0,-k_{1}-j+p+1;-%
\end{array}%
\right. \right) ,  \label{ri}
\end{align}%
$\Lambda _{1}\left( x,q,v\right) =\left(
\begin{array}{c}
-;g(x,0,v),1,2+q \\
g(x,q,v),1+q,0,-k_{1}-j+p+1;-%
\end{array}%
\right) $, and

$\Lambda _{2}\left( x,q,v\right) =\left(
\begin{array}{c}
-;g(x,0,v),1 \\
0,-k_{1}-j+p+1,g(x,q,v);-%
\end{array}%
\right) $%
\begin{IEEEproof}
The proof is provided in Appendix E.
\end{IEEEproof}
\end{proposition}

\subsection{Asymptotic Analysis}

\begin{proposition}
The system's IP\ can be asymptotically expanded at high SNR\ regime (i.e., $%
\overline{\gamma }_{\Xi }^{(l_{i})},\mu _{1}\rightarrow \infty )$ as follows%
\begin{equation}
P_{int,\Xi }^{\left( i,\infty \right) }=G_{c,\Xi }\left( \overline{\gamma }%
_{\Xi }^{(l_{i})}\right) ^{-G_{d}},
\end{equation}%
with%
\begin{align}
G_{c,\Xi }& =\left\{
\begin{array}{l}
\left( a_{0}^{(d)}\mathcal{P}_{1}\left( \epsilon _{\Xi }^{(i)}\Upsilon
^{\prime }a^{(\Xi )}\right) ^{x_{d}}\right) ^{K}F_{\gamma _{1}^{(e)}}\left(
a^{(\Xi )}\right)  \\
+\mathcal{X}\left( a^{(\Xi )}\right) ,\text{ if }Kx_{d}<1\text{ } \\
\mathcal{Q}^{(i,\Xi )},\text{ }Kx_{d}>1%
\end{array}%
\right. ,  \label{gc} \\
a^{(\Xi )}& =\left\{
\begin{array}{c}
\gamma _{th},\text{ }\Xi =NZF\text{ or }\left( \Xi =ZF\text{ and }\gamma
_{th}\geq \mathscr{L}_{i}\right)  \\
\mathscr{L}_{i},\text{ }\Xi =ZF\text{ and }\gamma _{th}<\mathscr{L}_{i}%
\end{array}%
\right. ,  \label{aa} \\
G_{d}& =\min \left( 1,Kx_{d}\right) ,  \label{gd}
\end{align}%
where $\Upsilon ^{\prime }=\frac{\xi _{1}^{2}\alpha _{1}\beta _{1}}{\xi
_{1}^{2}+1},$ $\epsilon _{\Xi }^{(i)}=\frac{\overline{\gamma }_{\Xi
}^{(l_{i})}}{\mu _{1}},$ $x_{d}=\underset{i=1,2,3}{\min }\left( x_{i}\right)
,$ and%
\begin{align}
\mathcal{X}\left( \varphi \right) & =\mathcal{P}_{e}\left( a_{0}^{(d)}%
\mathcal{P}_{1}\left( \epsilon _{\Xi }^{(i)}\Upsilon ^{\prime }\right)
^{x_{d}}\right) ^{K}  \notag \\
& \times \left[
\begin{array}{l}
\frac{\Gamma \left( \alpha _{e}+Kx_{d}\right) \Gamma \left( \beta
_{e}+Kx_{d}\right) }{\Upsilon _{e}^{Kx_{d}}\left( \xi _{e}^{2}+Kx_{d}\right)
} \\
-\varphi ^{Kx_{d}}G_{2,4}^{3,1}\left( \Upsilon _{e}\varphi \left\vert
\begin{array}{c}
1-Kx_{d};\xi _{e}^{2}+1 \\
\xi _{e}^{2},\alpha _{e},\beta _{e};-Kx_{d}%
\end{array}%
\right. \right)
\end{array}%
\right] .  \label{Xt}
\end{align}%
and $\mathcal{Q}^{(i,ZF)}$, $\mathcal{Q}^{(i,NZF)}$ are given in (\ref{qzf})
and (\ref{qnzf}) at the top of the next page, with $\mathcal{Z}^{(i)}=\frac{%
v_{l_{i}}\Psi ^{\left( e_{i}^{(2)}\right) }}{\mathcal{T}_{i}\exp \left(
\mathcal{G}^{\left( e_{i}^{(2)},l_{i}\right) }\right) },$ $E_{i}(.)$ is the
exponential integral function \cite[Eq. (8.211.1)]{integrals}.
\begin{figure*}[t]
{\normalsize 
\setcounter{mytempeqncnt}{\value{equation}}
\setcounter{equation}{47} }%
\begin{align}
Q^{(i,ZF)}& =\psi _{i}\exp \left( r_{1}^{\left( e_{i}^{(2)}\right) }\right)
\sum\limits_{n_{2}=0}^{m_{s}^{\left( e_{i}^{(2)}\right) }-1}U_{i}\left(
0,n_{2}\right) \sum\limits_{p=0}^{n_{2}+1}\binom{n_{2}+1}{p}%
(-1)^{n_{2}+1-p}\left( r_{1}^{\left( e_{i}^{(2)}\right) }\right)
^{n_{2}-p+1}\Gamma \left( p,r_{1}^{\left( e_{i}^{(2)}\right) }\right)
\label{qzf} \\
& \mathcal{Q}^{(i,NZF)}=\left\{
\begin{array}{l}
\mathcal{Z}^{(i)}\sum\limits_{n_{2}=0}^{m_{s}^{\left( e_{i}^{(2)}\right) }-1}%
\frac{\mathcal{U}_{i}\left( 0,n_{2}\right) }{\left( \mathcal{G}^{\left(
e_{i}^{(2)},l_{i}\right) }\right) ^{-n_{2}-1}}\left(
\begin{array}{l}
E_{i}\left( \mathcal{G}^{\left( e_{i}^{(2)},l_{i}\right) }\right)  \\
+\sum\limits_{q=1}^{n_{2}+1}\left(
\begin{array}{c}
n_{2}+1 \\
q%
\end{array}%
\right) \frac{\gamma _{inc}\left( q,-\mathcal{G}^{\left(
e_{i}^{(2)},l_{i}\right) }\right) }{\left( -\Theta ^{(l_{i})}\Psi ^{\left(
e_{i}^{(2)}\right) }\right) ^{-q-n_{2}-1}}%
\end{array}%
\right) ,\mathcal{L}^{(l_{i})}<\mathcal{L}^{\left( e_{i}^{(2)}\right) } \\
-\mathcal{Z}^{(i)}\sum\limits_{n_{2}=0}^{m_{s}^{\left( e_{i}^{(2)}\right)
}-1}\mathcal{U}_{i}\left( 0,n_{2}\right) \left( n_{2}+1\right)
!G_{1,2}^{2,0}\left( -\mathcal{G}^{\left( e_{i}^{(2)},l_{i}\right)
}\left\vert
\begin{array}{c}
2+n_{2} \\
0,1+n_{2}%
\end{array}%
\right. \right) ,\mathcal{L}^{(l_{i})}\geq \mathcal{L}^{\left(
e_{i}^{(2)}\right) }%
\end{array}%
,\right.   \label{qnzf}
\end{align}%
\par
{\normalsize 
\hrulefill 
\vspace*{4pt} }
\end{figure*}
\begin{IEEEproof}
The proof is provided in Appendix F.
\end{IEEEproof}
\end{proposition}

\section{Numerical results}

The derived analytic results, which are validated with their respective
Monte Carlo simulations are depicted in this section to analyze the
performance of the considered set up. Some illustrative numerical examples
are depicted to inspect the effects of the key system parameters on the
overall secrecy performance of the considered hybrid terrestrial-satellite
link. To this end, the system and channel parameters are set as detailed in
Table \ref{t1111}$.$ The FSO\ hop turbulence parameters were computed based
on OGS-satellite distance, wavelength, and aperture radius using \cite[Eq.
(4, 9-10)]{uplink}, \cite{farid}, \cite[Eq. (2)]{rubiosat}, and \cite[Eq. (8)%
]{sandalidis}. Furthermore, the positions of the legitimate and wiretap
nodes were set using Table I and based on (\ref{phi1}) with \ $\phi
_{j}^{(l_{j})}=3\times 10^{-3}$ rad$,$ $\phi _{j}^{\left( e_{j}^{(2)}\right)
}=6.66\times 10^{-4}$ rad , respectively, except Figs. 7-11, where $\phi
_{j}^{\left( e_{j}^{(2)}\right) }=6.66\times 10^{-4}$ rad$,$ $\phi
_{j}^{(l_{j})}=5\times 10^{-4}$ rad. Also, the simulation is performed by
generating $2\times 10^{6}$ random samples.
\begin{table*}[t]
\caption{{\protect\footnotesize Simulation {p}arameters' {v}alues.}}
\label{t1111}\centering
\par
\begin{tabular}{c|c||c|c||c|c}
\hline
Parameter & Value & Parameter & Value & Parameter & Value \\ \hline
$f$ & 20 GHz & $K$ & 3 (Except Fig. \ref{fig3sat}) & $N$ & 5 \\ \hline
$T$ & 193.15 $%
{{}^\circ}%
K$ & $R_{k}$ & 50 cm & $r$ & 6000 km \\ \hline
$B_{w}$ & 50 MHz & $R_{e}$ & 100 cm & $V_{w}$ & 21 m/s \\ \hline
$G_{R}^{\left( l_{i}\right) }$ & 40 dBi (except Fig. \ref{fig4sat}) & $%
\omega _{e}$ & 0.3 (Except Fig. \ref{fig6sat}) & $\frac{P_{S}}{\sigma
_{1}^{2}}$ & $80$ dB (Except Fig. \ref{fig6sat}) \\ \hline
$G_{R}^{\left( e_{i}^{(2)}\right) }$ & 25 dBi & $\omega _{l}$ & 0.7 (Except
Fig. \ref{fig6sat}) & $\frac{P_{S}}{\sigma _{e^{(1)}}^{2}}$ & $60$ dB
(Except Fig. \ref{fig6sat}) \\ \hline
$G_{T}^{(i)}$ & 60 dBi & $m_{s}^{(\varkappa _{i})}$ & 2 & $W_{0}$ & 8 cm \\
\hline
$\phi _{j,\text{3dB}}$ & 0.4$%
{{}^\circ}%
$ & $\Omega _{s}^{(\varkappa _{i})}$ & 3 (Except Fig. \ref{fig5sat}) & $%
\lambda $ & 1550 nm \\ \hline
$d_{k}=d_{e}$ & 6000 Km & $b_{\varkappa _{i}}$ & 1.4 (Except Fig. \ref%
{fig5sat}) & $D$ & 100 Km \\ \hline
$h_{0}$ & 25 m & $A$ & $1.7\times 10^{-14}$ m$^{-\frac{2}{3}}$ & $\gamma
_{th}$ & 20 dB \\ \hline
\end{tabular}%
\end{table*}
\begin{figure}[h]
\centering%
\begin{minipage}[t]{0.4\textwidth}
\begin{center}
\hspace*{-.8cm}\vspace*{-.4cm}\includegraphics[scale=.66]{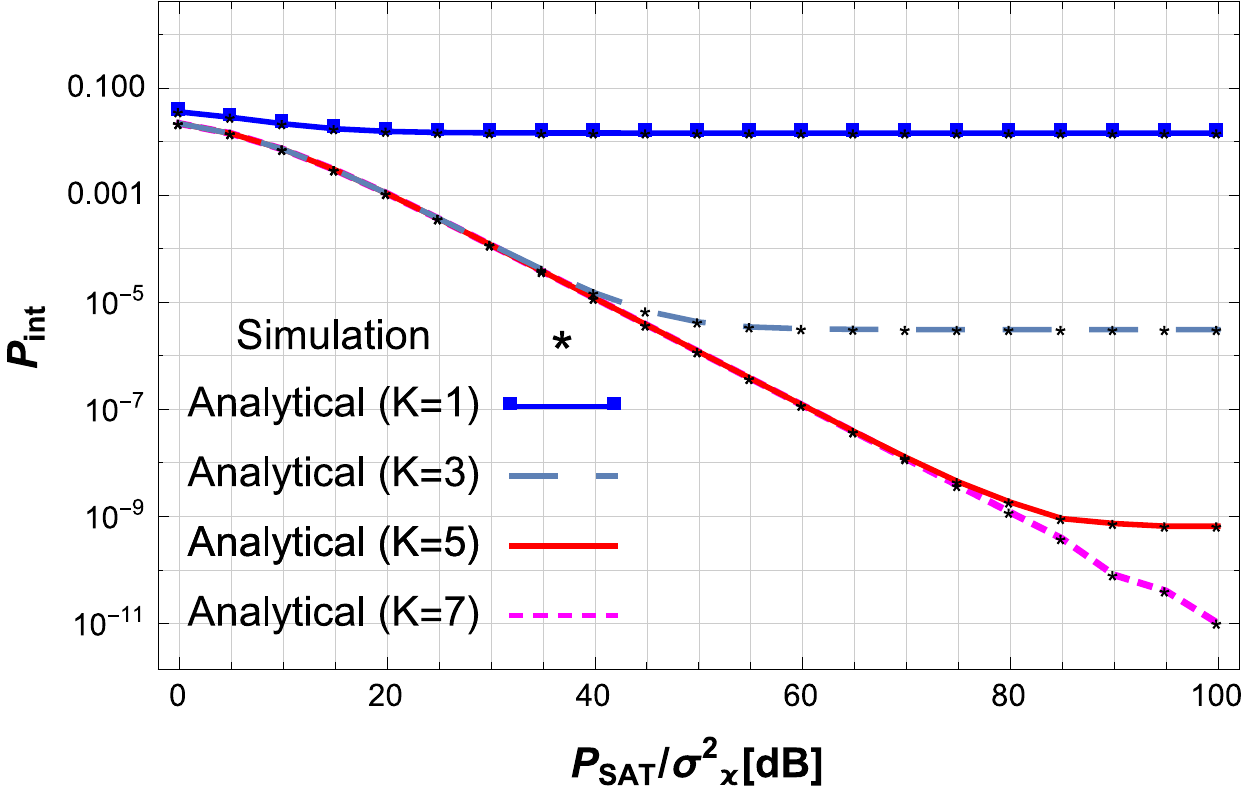}
\end{center}
\caption{{\footnotesize  IP versus $\frac{P_{SAT}}{\sigma %
_{\varkappa }^{2}}$ for various $K$ values.}}
\label{fig3sat}
  \end{minipage}\hfill
\begin{minipage}[t]{0.4\textwidth}
 \begin{center}
\hspace*{-.8cm}\vspace*{-.4cm}\includegraphics[scale=.66]{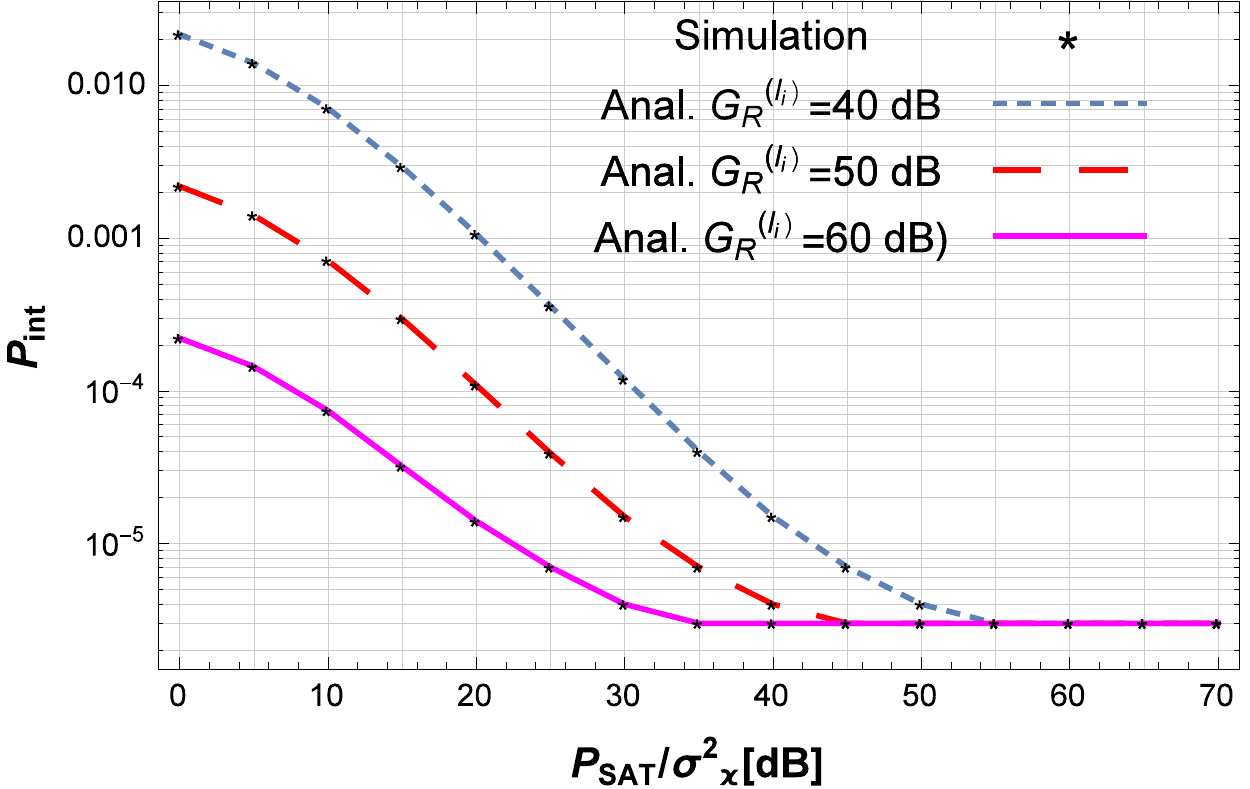}
\end{center}
\caption{{\footnotesize  IP versus $\frac{P_{SAT}}{\sigma %
_{\varkappa }^{2}}$ for various $G_R^{(l_i)}$ values.}}
\label{fig4sat}
  \end{minipage}
\end{figure}
\begin{figure}[t]
\centering\vspace*{-0.2cm}
\begin{minipage}[t]{0.4\textwidth}
\begin{center}
\hspace*{-.8cm}\vspace*{-.4cm}\includegraphics[scale=.66]{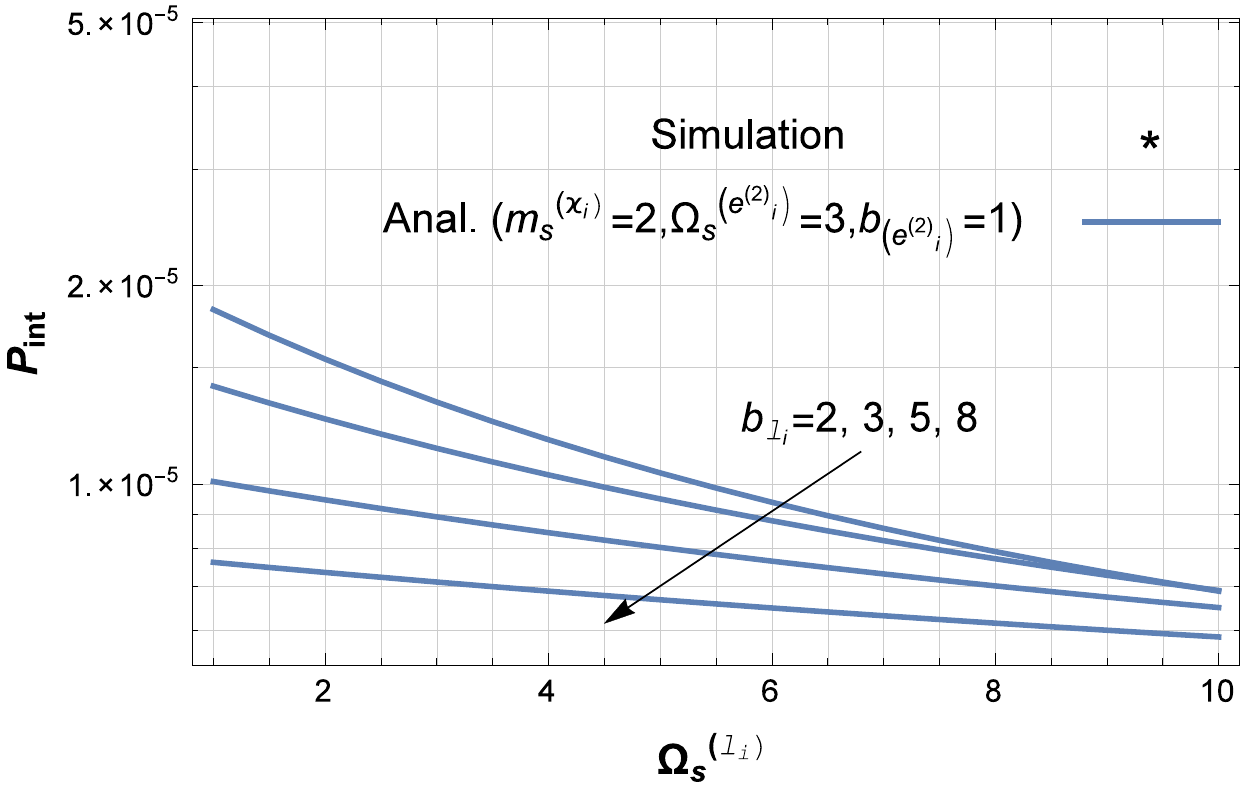}
\end{center}
\caption{{\footnotesize  IP without ZF\ precoding versus $\Omega _{s}^{(l_{i})}$}}
\label{fig5sat}
  \end{minipage}\hfill
\begin{minipage}[t]{0.4\textwidth}
\begin{center}
\hspace*{-.8cm}\vspace*{-.4cm}\includegraphics[scale=.66]{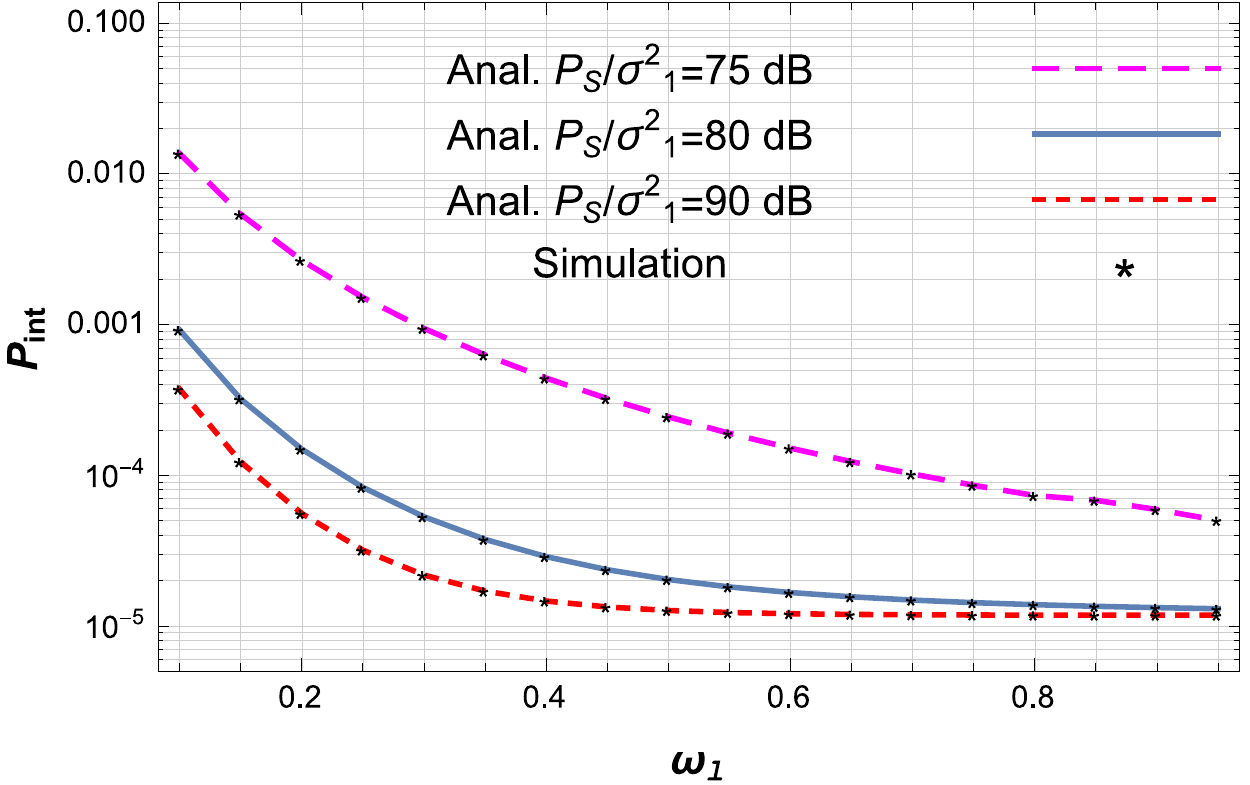}
\end{center}
\caption{{\footnotesize  IP with ZF\ precoding versus $\omega _{l}$ .}}
\label{fig6sat}
  \end{minipage}
\end{figure}
\begin{figure}[t]
\centering\vspace*{-0.2cm}
\begin{minipage}[t]{0.4\textwidth}
\begin{center}
\hspace*{-.8cm}\vspace*{-.4cm}\includegraphics[scale=.66]{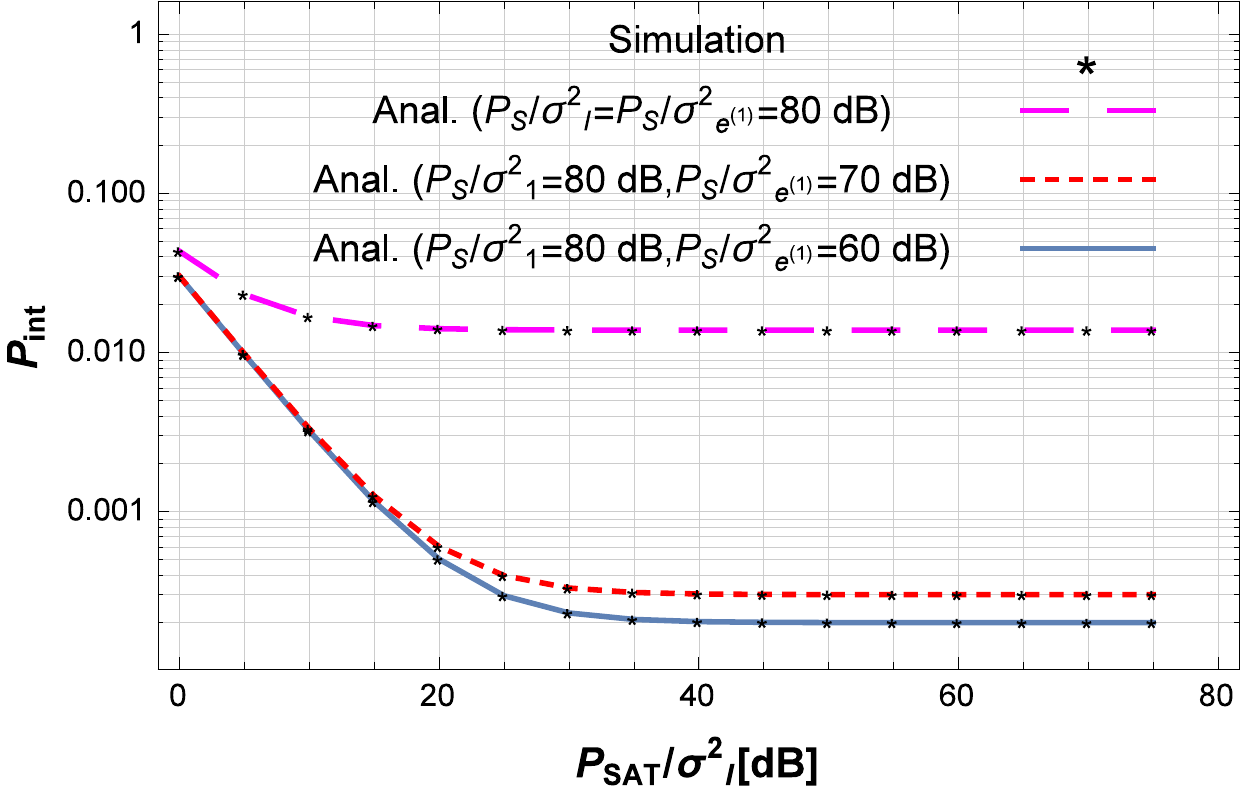}
\end{center}
\caption{{\footnotesize  IP with ZF precoding vs $\frac{P_{SAT}}{\sigma_l^2}.$}}
\label{figzf}
  \end{minipage}\hfill
\begin{minipage}[t]{0.4\textwidth}
\begin{center}
\hspace*{-.8cm}\vspace*{-.4cm}\includegraphics[scale=.66]{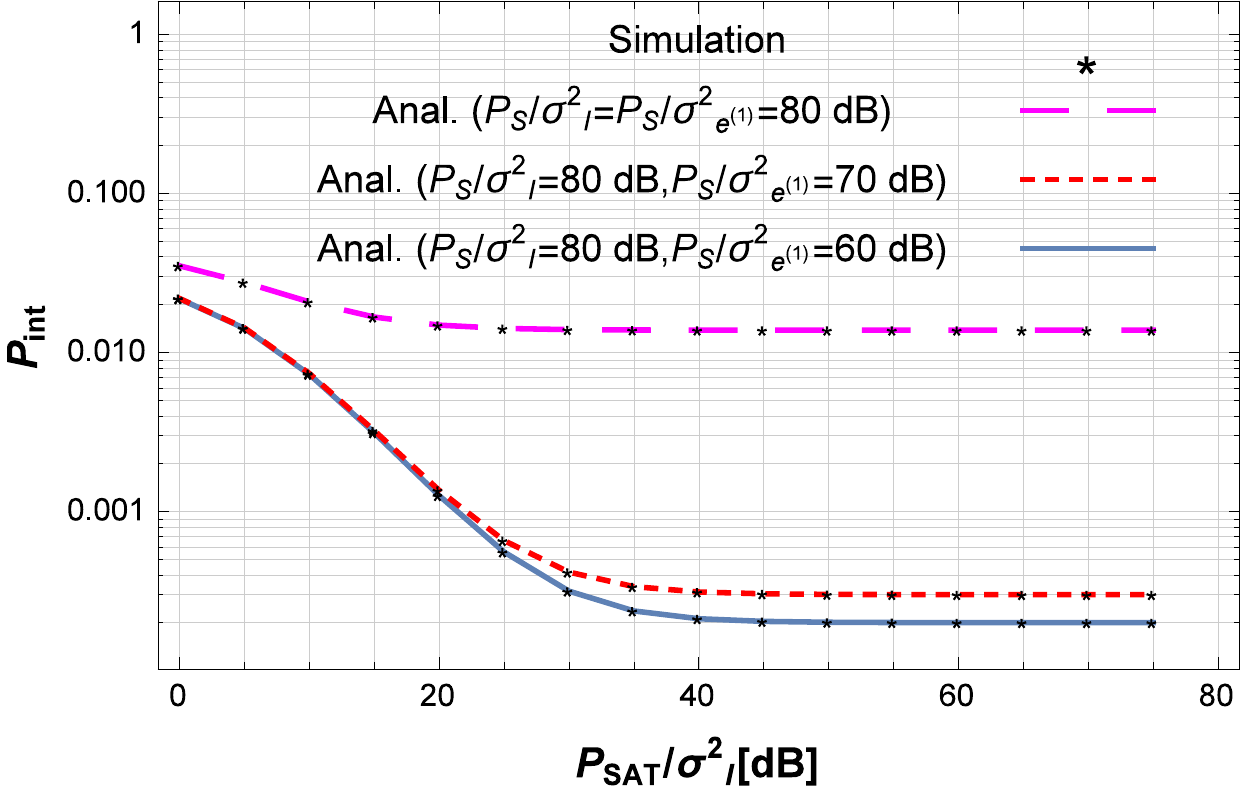}
\end{center}
\caption{{\footnotesize  IP without ZF\ precoding vs $\frac{P_{SAT}}{\sigma_l^2}$}}
\label{fignzf}
  \end{minipage}
\end{figure}
\begin{figure}[t]
\centering\vspace*{-0.2cm}
\begin{minipage}[t]{0.4\textwidth}
\begin{center}
\hspace*{-.8cm}\vspace*{-.4cm}\includegraphics[scale=.66]{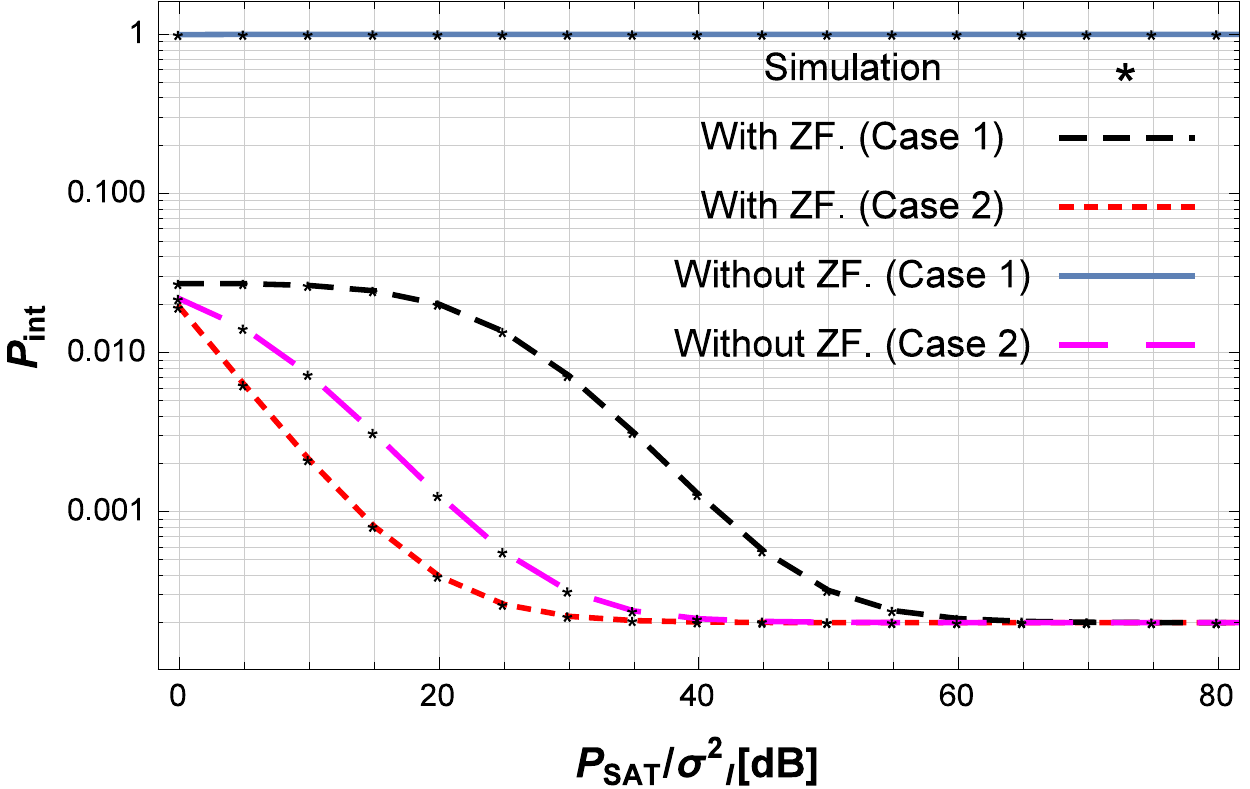}
\end{center}
\caption{{\footnotesize Comparison of the IP\ of both scenarios: With and without ZF.}}
\label{fig8}
  \end{minipage}\hfill
\begin{minipage}[t]{0.4\textwidth}
\begin{center}
\hspace*{-.8cm}\vspace*{-.4cm}\includegraphics[scale=.66]{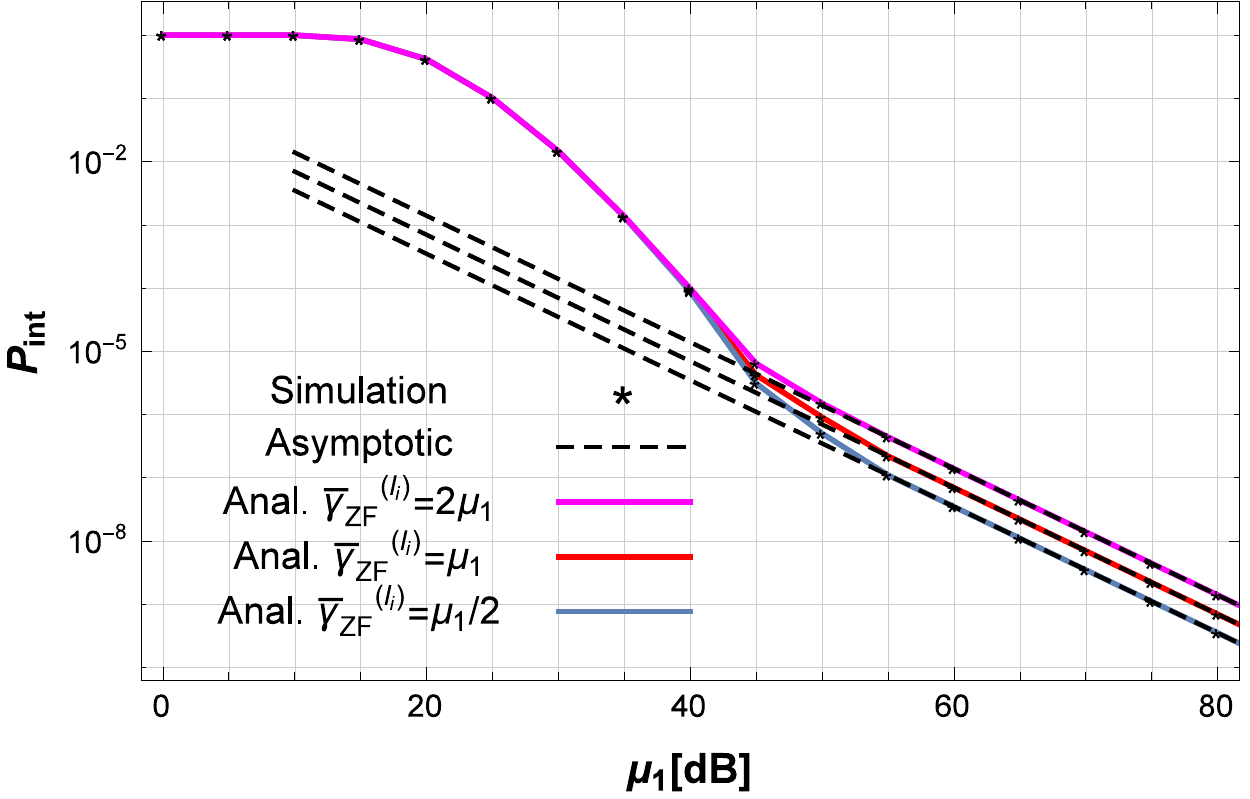}
\end{center}
\caption{{\footnotesize IP with ZF precoding vs $\mu_1$.}}
\label{figasm}
  \end{minipage}
\end{figure}

In Fig. \ref{fig3sat}, the IP\ is shown as a function of the average SNR\ $%
\frac{P_{SAT}}{\sigma _{\varkappa }^{2}}$ ($\varkappa =\{l,e^{(2)}\}),$ by
considering $\frac{P_{SAT}}{\sigma _{l}^{2}}=\frac{P_{SAT}}{\sigma
_{e^{(2)}}^{2}},$ for various optical apertures at the satellite. One can
ascertain evidently that the secrecy performance decreases as a function of $%
\frac{P_{SAT}}{\sigma _{\varkappa }^{2}}$, particularly below a certain
threshold SNR and for high values of $K$. However, the secrecy gets steady
for higher average SNR\ values. Also, the number of optical apertures at the
satellite impacts the secrecy level of the system. The greater the number of
apertures, the higher is the legitimate link capacity, and consequently, the
more secure is the whole link. Nevertheless, one can note also that above $%
K=7$ apertures, the IP\ is not improved significantly.

Fig. \ref{fig4sat} shows the IP\ evolution for the ZF\ case versus $\frac{%
P_{SAT}}{\sigma _{\varkappa }^{2}}$ for various values of legitimate
receiver antenna gains. One can remark that for increasing the antennas
gains of the legitimate users can ensure a significant improvement of the
received legitimate SNR, and consequently results in better secrecy
performance of the system.

Fig. \ref{fig5sat} depicts the IP\ for the non-ZF\ case as a function of $%
\Omega _{s}^{(l_{i})}$\ for various values $b_{l_{i}}$, with $\Omega
_{s}^{\left( e_{i}^{(2)}\right) }=3,$ $m_{s}^{(x_{i})}=2,$ $%
b_{e_{i}^{(2)}}=1 $, and $\frac{P_{SAT}}{\sigma _{l}^{2}}=\frac{P_{SAT}}{%
\sigma _{e^{(2)}}^{2}}=40$ dB. One can ascertain that the higher the power
of the legitimate link LOS and multipath components $\Omega _{s}^{(l_{i})}$
and $b_{l_{i}},$ the higher the second hop's legitimate link SNR, which
results in a greater overall secrecy capacity. Thus, the system's secrecy
gets better.

Fig. \ref{fig6sat} shows the IP evolution versus the power portion received
at the satellite $\omega _{l},$ for a fixed transmit power-to-noise ratio of
the second hop $\frac{P_{SAT}}{\sigma _{l}^{2}}=\frac{P_{SAT}}{\sigma
_{e^{(2)}}^{2}}=40$ dB, and $\frac{P_{S}}{\sigma _{1}^{2}}=50,$ $60$, and $%
70 $ dB with $\frac{P_{S}}{\sigma _{1}^{2}}=\frac{P_{S}}{\sigma
_{e^{(1)}}^{2}}+20$ dB. One can notice that the system's secrecy improves by
increasing the portion of power received by the legitimate node (i.e., the
satellite). In fact, as $\omega _{l}$ is increasing, the SNR\ of the first
hop's legitimate link as well as the equivalent end-to-end SNR increase. As
a result, the overall secrecy capacity in (\ref{cs1}) gets greater too,
which leads to a decrease in the system's IP. Nevertheless, at higher $%
\omega _{l}$ values (higher $\gamma _{1}$), the IP gets steady at a certain
level. In fact, from (\ref{cs1}), the overall secrecy capacity will be
restricted to the minimum of the two hops' secrecy capacities (i.e., $%
C_{s,eq,\Xi }^{(2,i)})$, regardless the increase in $\omega _{l}$ (i.e., $%
C_{s}^{(1)}$ increase).

\bigskip Figs. \ref{figzf} and \ref{fignzf} depict the IP\ evolution versus $%
\frac{P_{SAT}}{\sigma _{l}^{2}},$ for the ZF and the non-ZF\ precoding
scenarios, respectively, for three different scenarios of the first hop
SNRs, namely $\frac{P_{S}}{\sigma _{1}^{2}}=80$ dB, and $\frac{P_{S}}{\sigma
_{e^{(1)}}^{2}}=60,$ $70,$and $80$ dB. One can remarkably note that for the
both abovementioned scenarios, the IP\ is improved with increasing the
difference between the FSO\ legitimate and wiretapper average SNRs.\
Nevertheless, the greater the gap, the lesser the secrecy improvement.

In Fig. \ref{fig8}, the IP\ is shown versus $\frac{P_{SAT}}{\sigma _{l}^{2}}$
for two scenarios: (i) ZF\ precoding case, and (ii) non-ZF\ precoding. We
fixed again $\frac{P_{S}}{\sigma _{1}^{2}}=\frac{P_{S}}{\sigma _{e^{(1)}}^{2}%
}=30$ dB. Importantly, we depict the IP\ for the two abovementioned
scenarios, as computed in the non-ZF\ scenario in (\ref{ipclosedform2}), for
case 1 $\left( \mathcal{L}^{(l_{i})}<\mathcal{L}^{\left( e_{i}^{(2)}\right)
}\right) $ and case 2 $\left( \mathcal{L}^{(l_{i})}\geq \mathcal{L}^{\left(
e_{i}^{(2)}\right) }\right) $, based on Table I and (\ref{phi1}) with\ $\phi
_{j}^{\left( e_{j}^{(2)}\right) }=6.66\times 10^{-4}$ rad$,$ $\phi
_{j}^{(l_{j})}=5\times 10^{-4}$ rad $\left( \text{i.e., case 1}\right) $ and
\ $\phi _{j}^{(l_{j})}=3\times 10^{-3}$ rad, $\phi _{j}^{\left(
e_{j}^{(2)}\right) }=6.66\times 10^{-4}$ rad $\left( \text{i.e.,}\mathcal{\ }%
\text{case 2}\right) $, respectively. On the other hand, the two
aforementioned cases were implemented for the ZF\ precoding expression by
adopting the same two abovementioned scenarios of $\phi _{j}^{(\varkappa
_{i})}$. One can ascertain evidently that the ZF\ precoding case outperforms
its non-ZF\ counterpart for case 1$.$ Thus, the closer the legitimate node
to the beams boresight (cells' centers) compared to the wiretap node, the
worse is the secrecy. Additionally, the non-ZF\ scenario admits a steady IP\
behavior at high SNR\ regime regardless of the increase in $\frac{P_{SAT}}{%
\sigma _{l}^{2}}.$\ In fact, from (\ref{ti}), one can notice clearly that at
high $\overline{\gamma }_{NZF}^{(\varkappa _{i})}$ values, the received SNRs$%
\ \gamma _{\varkappa _{i},NZF}$ reduce to $\mathcal{L}^{(l_{i})}\ $and $%
\mathcal{L}^{\left( e_{i}^{(2)}\right) },$ for $\varkappa =l,e^{(2)},$
respectively. Given that $\mathcal{L}^{(l_{i})}<\mathcal{L}^{\left(
e_{i}^{(2)}\right) },$ it follows that the wiretapper received SNR\ is
always greater than the legitimate one for high average SNR\ values, leading
to steady IP\ performance. On the other hand, the IP\ shows a distinguished
behavior for the second case $\left( \text{i.e., }\phi _{j}^{(l_{i})}\geq
\phi _{j}^{\left( e_{i}^{(2)}\right) }\right) $, where the non-ZF\ case
outperforms the ZF\ precoding one. In fact, and similarly to the previous
case, as $\mathcal{L}^{(l_{i})}\geq \mathcal{L}^{\left( e_{i}^{(2)}\right) }$%
, the legitimate SNR\ is most likely greater than the wiretapper one. Thus,
the farther the legitimate node from the cells' center compared to $%
e_{i}^{(2)},$ the better the secrecy performance. Additionally, due to ZF\
SNR\ normalization in (\ref{phiii}) and (\ref{gambar2}), the ZF\ precoding
average SNR\ most likely drops below $\mathcal{L}^{(l_{i})}$, resulting in
secrecy performance degradation.

Fig. \ref{figasm} represents the IP evolution versus $\mu _{1}$,\ when ZF
precoding is performed, for various SNR$\ $proportionality values $\left(
\text{i.e., }\mu _{1}=\epsilon _{i}\overline{\gamma }_{ZF}^{\left(
l_{i}\right) },\text{ }\epsilon _{i}=\frac{1}{2},1,2\right) $, and for
eavesdropper average SNR\ value $\frac{P_{S}}{\sigma _{e^{(1)}}^{2}}=\frac{%
P_{S}}{\sigma _{e^{(2)}}^{2}}=30$ dB$.$ We can ascertain that the higher the
first hop average SNR (i.e., higher OGS\ transmit power or lower satellite
reception noise), the better the overall system's secrecy. Furthermore, the
asymptotic curves match tightly the exact ones, which proves the accuracy of
the retrieved expressions in high SNR\ regime.

\section{Conclusion}

The secrecy performance of a hybrid terrestrial-satellite communication
system, operating with an optical feeder in the presence of potential
wiretappers in both hops is assessed. Statistical properties of the per-hop
SNR\ of the legitimate, as well as the wiretap link, are derived.
Furthermore, a novel expression for the IP\ for a dual-hop DF-based
communication system is retrieved, based on which the analyzed system's IP
metric is investigated in closed-form and asymptotic expressions in terms of
key system and channel parameters for several parameters' values cases. The
analysis was performed for two scenarios, namely by considering ZF\
precoding at the satellite, and by assuming direct satellite delivery to the
earth-stations (i.e., non-ZF). The obtained analytical results show that the
system's secrecy can be significantly improved by increasing the number of
optical photodetectors at the satellite, gateway and satellite transmit
powers, as well as antennas gains. Furthermore,{\ ZF precoding technique
improves the system's secrecy level compared to the non-ZF\ case for some
specific nodes' positions scenarios.}

A potential extension of this work might be the consideration of
amplify-and-forward (AF) relaying scheme as well as investigating the impact
of cloud coverage on the overall secrecy performance.

\section*{Appendix A: Proof of Proposition 1}

The SNR\ $\gamma _{1}$ is the maximum among individual SNRs $\gamma
_{1}^{(k)}$ at each of the $K$ photodetectors at the satellite.
Consequently, its respective CDF\ can be expressed in the case of i.i.d
fading amplitudes, using (\ref{cdfx}) for coherent detection (i.e., $r=1$) as%
\begin{equation}
F_{\gamma _{1}}(z)=\left[ \mathcal{P}_{1}G_{2,4}^{3,1}\left( \Upsilon
z\left\vert
\begin{array}{c}
1;\xi _{1}^{2}+1 \\
\xi _{1}^{2},\alpha _{1},\beta _{1};0%
\end{array}%
\right. \right) \right] ^{K},
\end{equation}%
with $\Upsilon $ is defined in Proposition 1. By making use of the residues
theorem \cite[Theorem 1.2]{kilbas} on the Meijer's $G$-function above, one
obtains
\begin{equation}
F_{\gamma _{1}}(z)=\mathcal{P}_{1}^{K}\left[ \Delta _{1}+\Delta _{2}+\Delta
_{3}\right] ^{K},  \label{cdf11}
\end{equation}%
with
\begin{equation}
\Delta _{i}=\left( \Upsilon z\right) ^{x_{i}}\text{ }\mathcal{E}_{i},
\label{deltai}
\end{equation}%
and $\mathcal{E}_{i}=\sum_{l=0}^{\infty }a_{l}^{(i)}\left( \Upsilon z\right)
^{l},$ and $x_{i}\in \{\xi _{1}^{2},\alpha _{1},\beta _{1}\}.$

Using the multinomial theorem, the CDF\ can be formulated as%
\begin{equation}
F_{\gamma _{1}}(z)=\mathcal{P}_{1}^{K}\sum_{h_{1}+h_{2}+h_{3}=K}\frac{K!}{%
h_{1}!h_{2}!h_{3}!}\frac{\left( \Upsilon z\right) ^{h_{1}\xi
_{1}^{2}+h_{2}\alpha _{1}+h_{3}\beta _{1}}}{\mathcal{E}_{1}^{-h_{1}}\mathcal{%
E}_{2}^{-h_{2}}\mathcal{E}_{3}^{-h_{3}}}.  \label{stp2}
\end{equation}

Involving the identity \cite[Eq. (0.314)]{integrals}, we have the following:
$\mathcal{E}_{i}^{h_{i}}=\sum\limits_{l=0}^{\infty }c_{l}^{(i)}z^{l}\Upsilon
^{l},$ where the coefficients $c_{l}^{(i)}$ are defined in Proposition 1.
Consequently, we have the following
\begin{equation}
\mathcal{E}_{1}^{h_{1}}\mathcal{E}_{2}^{h_{2}}\mathcal{E}%
_{3}^{h_{3}}=c_{1}^{(1)}\sum\limits_{l=0}^{\infty }\left(
\sum\limits_{q_{2}+q_{3}=l}c_{q_{2}}^{(2)}c_{q_{3}}^{(3)}\right)
z^{l}\Upsilon ^{l}.  \label{prod}
\end{equation}

By involving (\ref{prod}) into (\ref{stp2}), Proposition 1 is attained.

\section*{Appendix B: Proof of Lemma 1}

\textbf{\ }Relying on the probability theory, we have the following%
\begin{equation}
\Pr \left( C_{s,\Xi }^{(i)}=0,\gamma _{1}>\gamma _{th}\right) =\Pr \left(
\gamma _{1}>\gamma _{th}\right) -\Pr \left( C_{s,\Xi }^{(i)}>0,\gamma
_{1}>\gamma _{th}\right) .  \label{pintident}
\end{equation}

When $\gamma _{1}<\gamma _{th},$ the satellite fails at decoding the
information message. Therefore, no signal will be transmitted to the
legitimate earth-stations $l_{i}$ as well as the eavesdroppers $e_{i}^{(2)}.$%
\ Hence, we have $\gamma _{l_{i},\Xi }=\gamma _{e_{i}^{(2)},\Xi }=0$, which
yields from {(\ref{cs2}) and (\ref{csmin}) that }$C_{s,\Xi }^{(i)}=0$ and $%
\Pr \left( \left. C_{s,\Xi }^{(i)}=0\right\vert \gamma _{1}<\gamma
_{th}\right) =1.$ As a result, by making use of {(\ref{pintident}) into (\ref%
{ipnew}),} the overall\ IP\ expression reduces to%
\begin{equation}
P_{int,\Xi }^{(i)}=1-\Pr \left( C_{s,\Xi }^{(i)}>0,\gamma _{1}>\gamma
_{th}\right) .
\end{equation}

By involving {(\ref{csmin}) into} the above equation and relying on
probability theory and some algebraic manipulations, one obtains%
\begin{equation}
P_{int,\Xi }^{(i)}=1-\underset{\mathcal{I}}{\underbrace{\Pr \left( \gamma
_{1}>\gamma _{1}^{(e)},\gamma _{1}>\gamma _{e_{i}^{(2)},\Xi },\gamma
_{l_{i},\Xi }>\gamma _{e_{i}^{(2)},\Xi },\gamma _{1}>\gamma _{th}\right) }}.
\end{equation}

The above probability $\mathcal{I}$ can be expressed as $\mathcal{I}%
=\sum\limits_{i=1}^{6}\mathcal{I}_{i},$ with $I_{i}$ are the probabilities
associated with the events $\boldsymbol{E}_{i}$ shown in Table II.
\begin{table}[tbp]
\caption{Six possible events for $\mathcal{I}$.}\centering
\begin{tabular}{c|c|c|c}
& Event &  & Event \\ \hline
$\boldsymbol{E}_{1}$ & $\gamma _{1}>\gamma _{1}^{(e)}>\gamma
_{e_{i}^{(2)},\Xi }>\gamma _{th}$ & $\boldsymbol{E}_{4}$ & $\gamma
_{1}>\gamma _{th}>\gamma _{1}^{(e)}>\gamma _{e_{i}^{(2)},\Xi }$ \\ \hline
$\boldsymbol{E}_{2}$ & $\gamma _{1}>\gamma _{e_{i}^{(2)},\Xi }>\gamma
_{1}^{(e)}>\gamma _{th}$ & $\boldsymbol{E}_{5}$ & $\gamma _{1}>\gamma
_{e_{i}^{(2)},\Xi }>\gamma _{th}>\gamma _{1}^{(e)}$ \\ \hline
$\boldsymbol{E}_{3}$ & $\gamma _{1}>\gamma _{th}>\gamma _{e_{i}^{(2)},\Xi
}>\gamma _{1}^{(e)}$ & $\boldsymbol{E}_{6}$ & $\gamma _{1}>\gamma
_{1}^{(e)}>\gamma _{th}>\gamma _{e_{i}^{(2)},\Xi }$ \\ \hline
\end{tabular}%
\end{table}

Relying on Table II, one can see that%
\begin{align}
\mathcal{I}_{1}& \mathcal{=}\int_{y=\gamma _{th}}^{\infty }F_{\gamma
_{1}}^{c}\left( y\right) f_{\gamma _{1}^{(e)}}\left( y\right)
dy\int_{z=\gamma _{th}}^{y}f_{\gamma _{e_{i}^{(2)},\Xi }}\left( z\right)
F_{\gamma _{l_{i},\Xi }}^{c}\left( z\right) dz,  \label{i1} \\
\mathcal{I}_{2}& =\int_{z=\gamma _{th}}^{\infty }F_{\gamma _{1}}^{c}\left(
z\right) f_{\gamma _{e_{i}^{(2)},\Xi }}\left( z\right) F_{\gamma _{l_{i},\Xi
}}^{c}\left( z\right) \int_{y=\gamma _{th}}^{z}f_{\gamma _{1}^{(e)}}\left(
y\right) dydz,  \label{i2} \\
\mathcal{I}_{3}& =\int_{x=\gamma _{th}}^{\infty }\int_{y=0}^{\gamma
_{th}}\int_{z=y}^{\gamma _{th}}\int_{t=z}^{\infty }f_{\gamma _{1}}\left(
x\right) f_{\gamma _{1}^{(e)}}\left( y\right) f_{\gamma _{e_{i}^{(2)},\Xi
}}\left( z\right)  \label{i3} \\
& \times f_{\gamma _{l_{i},\Xi }}\left( t\right) dxdydzdt,  \notag \\
\mathcal{I}_{4}& =\int_{x=\gamma _{th}}^{\infty }\int_{z=0}^{\gamma
_{th}}\int_{y=z}^{\gamma _{th}}\int_{t=z}^{\infty }f_{\gamma _{1}}\left(
x\right) f_{\gamma _{1}^{(e)}}\left( y\right) f_{\gamma _{e_{i}^{(2)},\Xi
}}\left( z\right)  \label{i4} \\
& \times f_{\gamma _{l_{i},\Xi }}\left( t\right) dxdydzdt,  \notag \\
\mathcal{I}_{5}& =\int_{y=0}^{\gamma _{th}}\int_{x=\gamma _{th}}^{\infty
}\int_{z=\gamma _{th}}^{x}\int_{t=z}^{\infty }f_{\gamma _{1}}\left( x\right)
f_{\gamma _{1}^{(e)}}\left( y\right) f_{\gamma _{e_{i}^{(2)},\Xi }}\left(
z\right)  \label{i5} \\
& \times f_{\gamma _{2,\Xi }^{(l_{i})}}\left( t\right) dxdydzdt,  \notag
\end{align}%
\begin{align}
\mathcal{I}_{6}& \mathcal{=}\int_{x=\gamma _{th}}^{\infty }\int_{y=\gamma
_{th}}^{x}\int_{z=0}^{\gamma _{th}}\int_{t=z}^{\infty }f_{\gamma _{1}}\left(
x\right) f_{\gamma _{1}^{(e)}}\left( y\right) f_{\gamma _{e_{i}^{(2)},\Xi
}}\left( z\right)  \notag \\
& \times f_{\gamma _{l_{i},\Xi }}\left( t\right) dxdydzdt.  \label{i6}
\end{align}

By applying an integration by parts on ($\mathcal{\ref{i1}}$) and ($\mathcal{%
\ref{i3}}$) with $u^{\prime }(y)=f_{\gamma _{1}^{(e)}}\left( y\right) ,$ and
on ($\mathcal{\ref{i5}}$) with $u^{\prime }(x)=f_{\gamma _{1}}\left(
x\right) $ alongside with some algebraic manipulations, one obtains%
\begin{align}
\mathcal{I}_{1}& =\int_{y=\gamma _{th}}^{\infty }f_{\gamma _{1}}\left(
y\right) F_{\gamma _{1}^{(e)}}\left( y\right) dy\int_{z=\gamma
_{th}}^{y}f_{\gamma _{e_{i}^{(2)},\Xi }}\left( z\right) F_{\gamma
_{l_{i},\Xi }}^{c}\left( z\right) dz  \notag \\
& -\int_{y=\gamma _{th}}^{\infty }F_{\gamma _{1}}^{c}\left( y\right)
f_{\gamma _{e_{i}^{(2)},\Xi }}\left( y\right) F_{\gamma _{l_{i},\Xi
}}^{c}\left( y\right) F_{\gamma _{1}^{(e)}}\left( y\right) dy.
\end{align}%
\begin{equation}
\mathcal{I}_{3}=F_{\gamma _{1}}^{c}\left( \gamma _{th}\right)
\int_{y=0}^{\gamma _{th}}F_{\gamma _{1}^{(e)}}\left( y\right) f_{\gamma
_{e_{i}^{(2)},\Xi }}\left( y\right) F_{\gamma _{l_{i},\Xi }}^{c}\left(
y\right) dy.
\end{equation}%
\begin{equation}
\mathcal{I}_{5}=F_{\gamma _{1}^{(e)}}\left( \gamma _{th}\right) \left\{
\begin{array}{c}
\int_{z=\gamma _{th}}^{\infty }f_{\gamma _{e_{i}^{(2)},\Xi }}\left( z\right)
F_{\gamma _{2}}^{c}\left( z\right) dz \\
-\int_{x=\gamma _{th}}^{\infty }F_{\gamma _{1}}\left( x\right) f_{\gamma
_{e_{i}^{(2)},\Xi }}\left( x\right) F_{\gamma _{2}}^{c}\left( x\right) dx%
\end{array}%
\right\}
\end{equation}

\bigskip Additionally, by using the basic definition of the CDF in terms of
the respective PDF, it yields%
\begin{align}
\mathcal{I}_{2}& =\int_{z=\gamma _{th}}^{\infty }F_{\gamma _{1}}^{c}\left(
z\right) f_{\gamma _{e_{i}^{(2)},\Xi }}\left( z\right) F_{\gamma _{l_{i},\Xi
}}^{c}\left( z\right) F_{\gamma _{1}^{(e)}}\left( z\right) dz  \notag \\
& -F_{\gamma _{1}^{(e)}}\left( \gamma _{th}\right) \int_{z=\gamma
_{th}}^{\infty }F_{\gamma _{1}}^{c}\left( z\right) f_{\gamma
_{e_{i}^{(2)},\Xi }}\left( z\right) F_{\gamma _{l_{i},\Xi }}^{c}\left(
z\right) dz.
\end{align}

\begin{align}
\mathcal{I}_{4}& =F_{\gamma _{1}}^{c}\left( \gamma _{th}\right) F_{\gamma
_{1}^{(e)}}\left( \gamma _{th}\right) \int_{z=0}^{\gamma _{th}}f_{\gamma
_{e_{i}^{(2)},\Xi }}\left( z\right) F_{\gamma _{l_{i},\Xi }}^{c}\left(
z\right) dz  \notag \\
& -F_{\gamma _{1}}^{c}\left( \gamma _{th}\right) \int_{z=0}^{\gamma
_{th}}f_{\gamma _{e_{i}^{(2)},\Xi }}\left( z\right) F_{\gamma _{l_{i},\Xi
}}^{c}\left( z\right) F_{\gamma _{1}^{(e)}}\left( z\right) dz.
\end{align}

\begin{align}
\mathcal{I}_{6}& =\left[ \int_{x=\gamma _{th}}^{\infty }f_{\gamma
_{1}}\left( x\right) F_{\gamma _{1}^{(e)}}\left( x\right) dx\right]
\int_{z=0}^{\gamma _{th}}f_{\gamma _{e_{i}^{(2)},\Xi }}\left( z\right)
F_{\gamma _{l_{i},\Xi }}^{c}\left( z\right) dz  \notag \\
& -F_{\gamma _{1}}^{c}\left( \gamma _{th}\right) F_{\gamma _{1}^{(e)}}\left(
\gamma _{th}\right) \int_{z=0}^{\gamma _{th}}f_{\gamma _{e_{i}^{(2)},\Xi
}}\left( z\right) F_{\gamma _{l_{i},\Xi }}^{c}\left( z\right) dz.
\end{align}

By summing the terms $\mathcal{I}_{3},$ $\mathcal{I}_{4},$ and $\mathcal{I}%
_{6},$ one obtains%
\begin{align}
\mathcal{I}_{3}+\mathcal{I}_{4}+\mathcal{I}_{6}& =\left[ \int_{x=\gamma
_{th}}^{\infty }f_{\gamma _{1}}\left( x\right) F_{\gamma _{1}^{(e)}}\left(
x\right) dx\right]   \notag \\
& \times \left[ \int_{z=0}^{\gamma _{th}}f_{\gamma _{e_{i}^{(2)},\Xi
}}\left( z\right) F_{\gamma _{l_{i},\Xi }}^{c}\left( z\right) dz\right] .
\end{align}

In a similar manner, summing the terms $\mathcal{I}_{1},$ $\mathcal{I}_{2},$
and $\mathcal{I}_{5}$ gives the following%
\begin{align}
\mathcal{I}_{1}+\mathcal{I}_{2}+\mathcal{I}_{5}& =\int_{y=\gamma
_{th}}^{\infty }f_{\gamma _{1}}\left( y\right) F_{\gamma _{1}^{(e)}}\left(
y\right) dy\int_{z=\gamma _{th}}^{y}f_{\gamma _{e_{i}^{(2)},\Xi }}  \notag \\
& \times F_{\gamma _{l_{i},\Xi }}^{c}\left( z\right) dz.
\end{align}

Thus, by summing the abovementioned two formulas, (\ref{propip}) is achieved.

\section*{Appendix C: Proof of Lemma 2}

\subsection{ZF\ Case}

By involving (\ref{cdfgam2}) and (\ref{cdfeve2}) alongside with {{\cite[Eq.
(8.356.3)]{integrals} and \cite[Eq. (06.06.20.0003.01)]{wolfram}}} into $%
\mathcal{J}_{ZF}(y)$ in Lemma 1, one obtains%
\begin{align}
\mathcal{J}_{ZF}(y)& =F_{\gamma _{e_{i}^{(2)},ZF}}\left( y\right)
-\sum\limits_{n_{1}=0}^{m_{s}^{(l_{i})}-1}\sum\limits_{n_{2}=0}^{m_{s}^{%
\left( e_{i}^{(2)}\right) }-1}\frac{\mathcal{U}_{i}\left( n_{1},n_{2}\right)
}{n_{1}!}  \notag \\
& \times \left( \mathcal{Y}_{e_{i}^{(2)}}^{(ZF)}\right)
^{n_{2}+1}\int_{0}^{y}\frac{z^{n_{2}}\gamma _{inc}\left( n_{1}+1,\mathcal{Y}%
_{l_{i}}^{(ZF)}z\right) }{\exp \left( \frac{\mathcal{Y}_{e_{i}^{(2)}}^{(ZF)}z%
}{\psi _{i}-\theta _{i}z}\right) \left( \psi _{i}-\theta _{i}z\right)
^{n_{2}+2}}dz.
\end{align}%
with $\mathcal{Y}_{\varkappa _{i}}^{\left( \Xi \right) }=\frac{v_{\varkappa
_{i}}}{\overline{\gamma }_{\Xi }^{\left( \varkappa _{i}\right) }}$. By using
the lower-incomplete Gamma sum representation in \cite[Eq. (8.352.1)]%
{integrals} alongside with some algebraic manipulations, it yields%
\begin{align}
\mathcal{J}_{ZF}(y)&
=\sum\limits_{n_{1}=0}^{m_{s}^{(l_{i})}-1}\sum\limits_{n_{2}=0}^{m_{s}^{%
\left( e_{i}^{(2)}\right) }-1}\frac{\mathcal{U}_{i}\left( n_{1},n_{2}\right)
}{\left( \mathcal{Y}_{e_{i}^{(2)}}^{(ZF)}\right) ^{-n_{2}-1}}%
\sum_{k_{1}=0}^{n_{1}}\frac{\left( \mathcal{Y}_{l_{i}}^{(ZF)}\right) ^{k_{1}}%
}{k_{1}!}  \notag \\
& \times \int_{0}^{y}\frac{z^{n_{2}+k_{1}}\exp \left[ -\left( \frac{1}{%
\mathcal{Y}_{e_{i}^{(2)}}^{(ZF)}\left( \psi _{i}-\theta _{i}z\right) }+%
\mathcal{Y}_{l_{i}}^{(ZF)}\right) z\right] }{\left( \psi _{i}-\theta
_{i}z\right) ^{n_{2}+2}}dz.
\end{align}

Based on the change of variable $w=\frac{z}{\psi _{i}-\theta _{i}z}+\frac{1}{%
\theta _{i}},$ we have the following%
\begin{align}
\mathcal{J}_{ZF}(y)& =\frac{\exp \left( \mathcal{Y}_{e_{i}^{(2)}}^{(ZF)}%
\theta _{i}^{-1}\right) }{\psi _{i}}\sum%
\limits_{n_{1}=0}^{m_{s}^{(l_{i})}-1}\sum\limits_{n_{2}=0}^{m_{s}^{\left(
e_{i}^{(2)}\right) }-1}\frac{\mathcal{U}_{i}\left( n_{1},n_{2}\right) }{%
\left( \mathcal{Y}_{e_{i}^{(2)}}^{(ZF)}\right) ^{-n_{2}-1}}  \notag \\
& \times \sum_{k_{1}=0}^{n_{1}}\frac{(-1)^{k_{1}}}{k_{1}!}%
\sum\limits_{j=0}^{\infty }\frac{\left( \psi _{i}\mathcal{Y}%
_{l_{i}}^{(ZF)}\right) ^{j+k_{1}}}{j!}\sum\limits_{p=0}^{k_{1}+n_{2}+j}%
\binom{k_{1}+n_{2}+j}{p}  \notag \\
& \times \left( -\theta _{i}\right) ^{p-n_{2}}\int_{\frac{1}{\theta _{i}}}^{%
\frac{y}{\psi _{i}-\theta _{i}y}+\frac{1}{\theta _{i}}}w^{-k_{1}-j+p}\exp
\left( -\mathcal{Y}_{e_{i}^{(2)}}^{(ZF)}w\right) dw.
\end{align}

Finally, using the upper-incomplete Gamma function definition in \cite[Eq.
(8.350.2)]{integrals}, (\ref{lemma21}) is attained.

\subsection{Non-ZF\ Case}

By incorporating the CDF\ and PDF expressions from (\ref{cdfnzf}) jointly
with {{\cite[Eq. (8.356.3)]{integrals} and \cite[Eq. (06.06.20.0003.01)]%
{wolfram}}} into $\mathcal{J}_{NZF}(y)$ in\ Lemma 1, one obtains%
\begin{align}
\mathcal{J}_{NZF}(y)& =F_{\gamma _{e_{i}^{(2)},NZF}}\left( y\right) -\Psi
^{\left( e_{i}^{(2)}\right)
}\sum\limits_{n_{1}=0}^{m_{s}^{(l_{i})}-1}\sum\limits_{n_{2}=0}^{m_{s}^{%
\left( e_{i}^{(2)}\right) }-1}\frac{\mathcal{U}_{i}\left( n_{1},n_{2}\right)
}{\left( \mathcal{Y}_{e_{i}^{(2)}}^{(NZF)}\right) ^{-n_{2}-1}}  \notag \\
& \times \int_{z=0}^{y}\frac{z^{n_{2}}\exp \left( -\frac{\mathcal{Y}%
_{e_{i}^{(2)}}^{(ZF)}z}{\left( \Psi ^{\left( e_{i}^{(2)}\right) }-\Theta
^{\left( e_{i}^{(2)}\right) }z\right) }\right) }{\left( \Psi ^{\left(
e_{i}^{(2)}\right) }-\Theta ^{\left( e_{i}^{(2)}\right) }z\right) ^{n_{2}+2}}
\notag \\
& \times \gamma _{inc}\left( n_{1}+1,\frac{\mathcal{Y}_{l_{i}}^{(NZF)}z}{%
\Psi ^{(l_{i})}-\Theta ^{(l_{i})}z}\right) dz.
\end{align}

By transforming the lower-incomplete Gamma function in {(\ref{cdfgam2})} to
an upper incomplete one alongside with the finite sum expression of the
upper-incomplete Gamma function {\cite[Eq. (8.352.2)]{integrals}, one can
see that}%
\begin{align}
\mathcal{J}_{NZF}(y)& =F_{\gamma _{e_{i}^{(2)},NZF}}\left( y\right)
-F_{\gamma _{e_{i}^{(2)},NZF}}\left( y\right) +\Psi ^{\left(
e_{i}^{(2)}\right) }  \notag \\
& \times
\sum\limits_{n_{1}=0}^{m_{s}^{(l_{i})}-1}\sum\limits_{n_{2}=0}^{m_{s}^{%
\left( e_{i}^{(2)}\right) }-1}\frac{\mathcal{U}_{i}\left( n_{1},n_{2}\right)
}{\left( \mathcal{Y}_{e_{i}^{(2)}}^{(NZF)}\right) ^{-n_{2}-1}}%
\sum\limits_{k_{1}=0}^{n_{1}}\frac{\left( \mathcal{Y}_{l_{i}}^{(NZF)}\right)
^{k_{1}}}{k_{1}!}  \notag \\
& \times \int_{z=0}^{y}\frac{\exp \left[ -z\left( \frac{\mathcal{Y}%
_{l_{i}}^{(NZF)}}{\Psi ^{(l_{i})}-\Theta ^{(l_{i})}z}+\frac{\mathcal{Y}%
_{e_{i}^{(2)}}^{(NZF)}}{\Psi ^{\left( e_{i}^{(2)}\right) }-\Theta ^{\left(
e_{i}^{(2)}\right) }z}\right) \right] }{\left( \Psi ^{\left(
e_{i}^{(2)}\right) }-\Theta ^{\left( e_{i}^{(2)}\right) }z\right) ^{n_{2}+2}}
\notag \\
& \times \frac{z^{n_{2}+k_{1}}}{\left( \Psi ^{(l_{i})}-\Theta
^{(l_{i})}z\right) ^{k_{1}}}dz.
\end{align}

{At this level, }two subcases for the result are distinguished{, namely }$%
\mathcal{L}^{\left( e_{i}^{(2)}\right) }>\mathcal{L}^{(l_{i})},$ and $%
\mathcal{L}^{\left( e_{i}^{(2)}\right) }<\mathcal{L}^{(l_{i})}.$ {By making
use of the change of variable }$w=\left\{
\begin{array}{c}
\frac{\Psi ^{(l_{i})}\mathcal{T}_{i}}{\Psi ^{(l_{i})}-\Theta ^{(l_{i})}z}%
+\Psi ^{(l_{i})}\Theta ^{\left( e_{i}^{(2)}\right) },\text{ if }\mathcal{L}%
^{\left( e_{i}^{(2)}\right) }>\mathcal{L}^{(l_{i})} \\
\mathcal{-}\frac{\Psi ^{\left( e_{i}^{(2)}\right) }\mathcal{T}_{i}}{\Psi
^{\left( e_{i}^{(2)}\right) }-\Theta ^{\left( e_{i}^{(2)}\right) }z}+\Theta
^{(l_{i})}\Psi ^{\left( e_{i}^{(2)}\right) },\text{ if }\mathcal{L}^{\left(
e_{i}^{(2)}\right) }<\mathcal{L}^{(l_{i})}%
\end{array}%
\right. ${\ }alongside with {\cite[Eq. (1.211.1)]{integrals} and the}
binomial theorem, yields ({\ref{antesult}}) at the top of the next page.
Finally, by using {\cite[Eq. (8.350.2)]{integrals}, }$\mathcal{J}_{NZF}(y)$
in {(\ref{lemma22}) is attained. }
\begin{figure*}[t]
{\normalsize 
\setcounter{mytempeqncnt}{\value{equation}}
\setcounter{equation}{76} }
\par
\begin{align}
\mathcal{J}_{NZF}(y)&
=\sum\limits_{n_{1}=0}^{m_{s}^{(l_{i})}-1}\sum\limits_{n_{2}=0}^{m_{s}^{%
\left( e_{i}^{(2)}\right) }-1}\frac{\mathcal{U}_{i}\left( n_{1},n_{2}\right)
}{\exp \left( -\mathcal{G}^{\left( l_{i},e_{i}^{(2)}\right) }\right) }%
\sum\limits_{k_{1}=0}^{n_{1}}\frac{\left( \frac{\mathcal{Y}%
_{l_{i}}^{(NZF)}\Psi ^{\left( e_{i}^{(2)}\right) }}{\mathcal{T}_{i}}\right)
^{k_{1}}}{k_{1}!}\sum_{j=0}^{\infty }\frac{\left( -1\right) ^{j}}{j!}\left(
-\Theta ^{(l_{i})}\Psi ^{\left( e_{i}^{(2)}\right) }\mathcal{G}^{\left(
e_{i}^{(2)},l_{i}\right) }\right) ^{j+n_{2}+1}  \notag \\
& \times \sum\limits_{p=0}^{n_{2}+k_{1}+j}\frac{(-1)^{p+1}}{\left( \Theta
^{(l_{i})}\Psi ^{\left( e_{i}^{(2)}\right) }\right) ^{p}}\binom{n_{2}+k_{1}+j%
}{p}\int_{w=\Theta ^{(l_{i})}\Psi ^{\left( e_{i}^{(2)}\right) }}^{\frac{%
\mathcal{T}_{i}\Psi ^{(l_{i})}}{\Psi ^{(l_{i})}-\Theta ^{(l_{i})}y}+\Theta
^{\left( e_{i}^{(2)}\right) }\Psi ^{(l_{i})}}\frac{w^{p-n_{2}-2-j}}{\exp
\left( \frac{r_{1}^{l_{i}}w}{\mathcal{T}_{i}}\right) }dw.  \label{antesult}
\end{align}%
\par
{\normalsize 
\hrulefill 
\vspace*{4pt} }
\end{figure*}

\begin{remark}
For the first study subcase (i.e., $\mathcal{L}^{\left( e_{i}^{(2)}\right) }>%
\mathcal{L}^{(l_{i})})$, {and from the integral definition }$\mathcal{J}%
_{\Xi }(y)$ in Lemma 1{, we have the following}
\begin{align}
\mathcal{J}_{NZF}\left( \mathcal{L}^{\left( e_{i}^{(2)}\right) }\right) &
=\int_{0}^{\mathcal{L}^{(l_{i})}}f_{\gamma _{e_{i}^{(2)},NZF}}\left(
z\right) \left( 1-F_{\gamma _{e_{i}^{(2)},NZF}}\left( z\right) \right) dz
\notag \\
& +\int_{\mathcal{L}^{(l_{i})}}^{\mathcal{L}^{\left( e_{i}^{(2)}\right)
}}f_{\gamma _{e_{i}^{(2)},NZF}}\left( z\right) \underset{0}{\underbrace{%
F_{\gamma _{l_{i},NZF}}^{c}\left( y\right) }}dz  \notag \\
& =\mathcal{J}_{NZF}\left( \mathcal{L}^{(l_{i})}\right) .
\end{align}

Thus, when evaluating $\mathcal{J}_{NZF}\left( .\right) $ for $y=\mathcal{L}%
^{\left( e_{i}^{(2)}\right) }$ when $\mathcal{L}^{\left( e_{i}^{(2)}\right)
}>\mathcal{L}^{(l_{i})},$ it should be evaluated at $\mathcal{L}^{(l_{i})}$
within {the final result in (\ref{lemma22}) for the non-ZF\ case.}
\end{remark}

\section{Appendix D: Proof of Lemma 3}

Relying on integration by parts in {(\ref{kphi}), the integral }$K\left(
\varphi \right) $ can be expressed as%
\begin{equation}
K\left( \varphi \right) =1-\left[ F_{\gamma _{1}}\left( y\right) F_{\gamma
_{1}^{(e)}}\left( y\right) \right] _{\varphi }^{\infty }-\mathcal{H}_{1}+%
\mathcal{H}_{2}\left( \varphi \right) .  \label{kphip}
\end{equation}%
with $\mathcal{H}_{1}=\int_{y=0}^{\infty }F_{\gamma _{1}}\left( y\right)
f_{\gamma _{1}^{(e)}}\left( y\right) dy$ and $\mathcal{H}_{2}\left( \varphi
\right) =\int_{y=0}^{\varphi }F_{\gamma _{1}}\left( y\right) f_{\gamma
_{1}^{(e)}}\left( y\right) dy$. Involving the CDF\ and PDF\ expressions in {(%
\ref{pdfx}) with parameters }$r=1,$ $\alpha _{e},$ $\beta _{e},$ $\xi
_{e}^{2},$ and $\mu _{e},${\ and (\ref{cdfgam1}) into the abovementioned
equations, and making use of the Mellin transform \cite[Eq. (2.9)]{mathai},
the integral }$\mathcal{H}_{1}$ is computed as%
\begin{align}
\mathcal{H}_{1}& =\mathcal{P}_{e}\mathcal{P}_{1}^{K}%
\sum_{h_{1}+h_{2}+h_{3}=K}\sum\limits_{l=0}^{\infty }\mathcal{F}%
_{h_{1},h_{2},h_{3},l}  \notag \\
& \times \frac{\Gamma \left( \alpha _{e}+\varrho
_{l,h_{1},h_{2},h_{3}}\right) \Gamma \left( \beta _{e}+\varrho
_{l,h_{1},h_{2},h_{3}}\right) }{\Upsilon _{e}^{\varrho
_{l,h_{1},h_{2},h_{3}}}\left( \xi _{e}^{2}+\varrho
_{l,h_{1},h_{2},h_{3}}\right) }.
\end{align}

Furthermore, by using {\cite[Eq. (06.10.02.0001.01)]{wolfram}, the above
equation reduces to (\ref{H1}). }Again, by using {\cite[Eq.
(07.34.21.0003.01)]{wolfram} in }$\mathcal{H}_{2}\left( \varphi \right) $, {(%
\ref{H2}) is attained.}

\section{Appendix E: Proof of Proposition 2}

\subsection{ZF\ Case}

By using Lemma 1 result as well as the PDF\ and CDF\ expressions of $\gamma
_{e_{i}^{(2)},ZF}$ and $\gamma _{l_{i},ZF}$ in {(\ref{cdfgam2}) and (\ref%
{cdfeve2}), we distinguish two cases, namely }$\mathscr{L} _{i}<\gamma
_{th}<y$ and $\mathscr{L} _{i}>\gamma _{th}.$

\begin{itemize}
\item \bigskip First case $\left( \mathscr{L}_{i}<\gamma _{th}\right) $: {as
the PDF\ }$f_{\gamma _{e_{i}^{(2)},ZF}}\left( z\right) =0$ for $z>\mathscr{L}%
_{i}$, the integral in {(\ref{propip}) becomes}%
\begin{equation}
\mathcal{I}=\underset{\mathcal{K}\left( \gamma _{th}\right) }{\underbrace{%
\left[ \int_{y=\gamma _{th}}^{\infty }f_{\gamma _{1}}\left( y\right)
F_{\gamma _{1}^{(e)}}\left( y\right) dy\right] }}\underset{\mathcal{J}%
_{ZF}\left( \mathscr{L}_{i}\right) }{\underbrace{\left[ \int_{z=0}^{%
\mathscr{L}_{i}}f_{\gamma _{e_{i}^{(2)},ZF}}\left( z\right) F_{\gamma
_{l_{i},ZF}}^{c}\left( z\right) dz\right] }},  \label{expI}
\end{equation}%
for $\mathscr{L}_{i}<\gamma _{th}.$ Thus, using Lemma 2 result for $y=%
\mathscr{L}_{i}$ as well as Lemma 3 result with $\varphi =\gamma _{th}$ and
involving it into {(\ref{expI}), one obtains}%
\begin{equation}
P_{int,ZF}^{(i)}=1-\mathcal{K}\left( \gamma _{th}\right) \mathcal{J}%
_{ZF}\left( \mathscr{L}_{i}\right) ;\mathscr{L}_{i}<\gamma _{th},
\end{equation}

\item Second case $\left( \mathscr{L}_{i}\geq \gamma _{th}\right) $: in such
an instance, two sub-cases are distinguished, namely $\mathscr{L}_{i}\geq
y\geq \gamma _{th}$ and $y>\mathscr{L}_{i}\geq \gamma _{th}$. Hence, the
integral {(\ref{propip}) becomes}
\begin{align}
\mathcal{I}& =\underset{\mathcal{O}^{(i)}}{\underbrace{\int_{y=\gamma
_{th}}^{\mathscr{L}_{i}}f_{\gamma _{1}}\left( y\right) F_{\gamma
_{1}^{(e)}}\left( y\right) \int_{z=0}^{y}f_{\gamma _{e_{i}^{(2)},ZF}}\left(
z\right) F_{\gamma _{l_{i},ZF}}^{c}\left( z\right) dzdy}}  \notag \\
& +\underset{\mathcal{N}^{(i)}}{\underbrace{\left[ \int_{y=\mathscr{L}%
_{i}}^{\infty }f_{\gamma _{1}}\left( y\right) F_{\gamma _{1}^{(e)}}\left(
y\right) dy\right] \left[ \int_{z=0}^{\mathscr{L}_{i}}f_{\gamma
_{e_{i}^{(2)},ZF}}\left( z\right) F_{\gamma _{l_{i},ZF}}^{c}\left( z\right)
dz\right] }}.
\end{align}
\end{itemize}

\bigskip Interestingly, the integrals of the second term $\mathcal{N}^{(i)}$
can be computed readily from Lemma 2 and Lemma 3, with $y=\mathscr{L}_{i}$
and $\varphi =\mathscr{L}_{i}$, respectively. On the other hand, by
involving Lemma 2 result in {(\ref{lemma21}) as well as the derivative of (%
\ref{cdfgam1}) and (\ref{pdfx}) with }$\left( \varpi =e{,\ }r=1\right) $
into $\mathcal{O}^{(i)}$, it produces the following%
\begin{eqnarray}
\mathcal{O}^{(i)} &\mathcal{=}&\int_{y=\gamma _{th}}^{\mathscr{L}%
_{i}}f_{\gamma _{1}}\left( y\right) F_{\gamma _{1}^{(e)}}\left( y\right)
\mathcal{J}(y)dy  \notag \\
&=&\mathcal{O}_{1}^{(i)}-\mathcal{O}\left( \mathscr{L}_{i}\right) .
\end{eqnarray}%
with $\mathcal{O}_{1}^{(i)}$ and $\mathcal{O}\left( \mathscr{L}_{i}\right) $
given in ({\ref{N1}}) and ({\ref{N2}}) at the top of the next page,
respectively.
\begin{figure*}[t]
{\normalsize 
\setcounter{mytempeqncnt}{\value{equation}}
\setcounter{equation}{84} }
\par
\begin{align}
\mathcal{O}_{1}^{(i)}& =\frac{\mathcal{P}_{e}\mathcal{P}_{1}^{K}\mathcal{Y}%
_{e_{i}^{(2)}}^{(ZF)}}{\exp \left( -r_{1}^{\left( e_{i}^{(2)}\right)
}\right) }\sum\limits_{n_{1}=0}^{m_{s}^{(l_{i})}-1}\sum%
\limits_{n_{2}=0}^{m_{s}^{\left( e_{i}^{(2)}\right) }-1}\mathcal{U}%
_{i}\left( n_{1},n_{2}\right)
\sum_{k_{1}=0}^{n_{1}}\sum\limits_{j=0}^{\infty }\frac{(-1)^{j}}{k_{1}!j!}%
\left( r_{1}^{(l_{i})}\psi _{i}\right)
^{j+k_{1}}\sum\limits_{p=0}^{h_{1}+n_{2}+j}\frac{\binom{h_{1}+n_{2}+j}{p}}{%
\left( -r_{1}^{\left( e_{i}^{(2)}\right) }\right) ^{p-h_{1}-j-n_{2}}}  \notag
\\
& \times \Gamma \left( -k_{1}-j+p+1,r_{1}^{\left( e_{i}^{(2)}\right)
}\right) \sum_{h_{1}+h_{2}+h_{3}=K}\sum\limits_{l=0}^{\infty }\mathcal{F}%
_{h_{1},h_{2},h_{3},l}\varrho _{l,h_{1},h_{2},h_{3}}\int_{y=\gamma _{th}}^{%
\mathscr{L}_{i}}\frac{G_{2,4}^{3,1}\left( \Upsilon _{e}y\left\vert
\begin{array}{c}
1;\xi _{e}^{2}+1 \\
\xi _{e}^{2},\alpha _{e},\beta _{e};0%
\end{array}%
\right. \right) }{y^{1-\varrho _{l,h_{1},h_{2},h_{3}}}}dy,  \label{N1}
\end{align}%
\par
{\normalsize 
\hrulefill 
\vspace*{4pt} }
\end{figure*}
\begin{figure*}[t]
{\normalsize 
\setcounter{mytempeqncnt}{\value{equation}}
\setcounter{equation}{85} }
\par
\begin{align}
\mathcal{O}\left( \mathscr{L}_{i}\right) & =\exp \left( r_{1}^{\left(
e_{i}^{(2)}\right) }\right) \mathcal{P}_{e}\mathcal{P}_{1}^{K}%
\sum_{h_{1}+h_{2}+h_{3}=K}\sum\limits_{l=0}^{\infty }\mathcal{F}%
_{h_{1},h_{2},h_{3},l}\varrho
_{l,h_{1},h_{2},h_{3}}\sum\limits_{n_{1}=0}^{m_{s}^{(l_{i})}-1}\sum%
\limits_{n_{2}=0}^{m_{s}^{\left( e_{i}^{(2)}\right) }-1}\mathcal{U}%
_{i}\left( n_{1},n_{2}\right)  \notag \\
& \times \sum_{h_{1}=0}^{n_{1}}\sum\limits_{j=0}^{\infty }\frac{%
(-1)^{j}\left( -r_{1}^{(l_{i})}\psi _{i}\right) ^{j}}{k_{1}!j!}%
\sum\limits_{p=0}^{h_{1}+n_{2}+j}\binom{h_{1}+n_{2}+j}{p}\left(
-r_{1}^{\left( e_{i}^{(2)}\right) }\right) ^{h_{1}+j-p+n_{2}}  \notag \\
& \times \underset{\mathcal{B}^{(1,i)}}{\underbrace{\int_{y=\gamma _{th}}^{%
\mathscr{L}_{i}}y^{\varrho _{l,h_{1},h_{2},h_{3}}-1}G_{2,4}^{3,1}\left(
\Upsilon _{e}y\left\vert
\begin{array}{c}
1;\xi _{e}^{2}+1 \\
\xi _{e}^{2},\alpha _{e},\beta _{e};0%
\end{array}%
\right. \right) \Gamma \left( -h_{1}-j+p+1,\mathcal{Y}_{e_{i}^{(2)}}^{(ZF)}%
\left( \frac{y}{\psi _{i}-\theta _{i}y}+\frac{1}{\theta _{i}}\right) \right)
dy}}.  \label{N2}
\end{align}%
\par
{\normalsize 
\hrulefill 
\vspace*{4pt} }
\end{figure*}

\bigskip Note that $\mathcal{O}\left( \mathscr{L}_{i}\right) $equals $%
\mathcal{O}_{2}^{(i)}$ or $\mathcal{O}_{3}^{(i)}$ given in {(\ref{Om})} for $%
\gamma _{th}\leq \mathscr{L}_{i}<2\gamma _{th}$ and $\mathscr{L}_{i}>2\gamma
_{th}$, respectively.

One can notice evidently {that (\ref{O1}) yields }from {(\ref{pdfx}), (\ref%
{cdfgam1}),} {(\ref{lemma21}), as well as (\ref{kphi}) and} {(\ref{N1})}. On
the other hand, the Meijer's $G$-function in {(\ref{N2}) can be expressed
using residues theorem as \cite[Theorem 1.2]{kilbas} }%
\begin{align}
G_{2,4}^{3,1}\left( \Upsilon _{e}y\left\vert
\begin{array}{c}
1;\xi _{e}^{2}+1 \\
\xi _{e}^{2},\alpha _{e},\beta _{e};0%
\end{array}%
\right. \right) & =b_{\xi _{e}^{2}}^{(0)}\left( \Upsilon _{e}y\right) ^{\xi
_{e}^{2}}+\sum_{v=0}^{\infty }b_{\alpha _{e}}^{(v)}\left( \Upsilon
_{e}y\right) ^{\alpha _{e}+v}  \notag \\
& +\sum_{v=0}^{\infty }b_{\beta _{e}}^{(v)}\left( \Upsilon _{e}y\right)
^{\beta _{e}+v}.  \label{residd}
\end{align}%
under the assumption: $x_{i}-x_{j}\notin
\mathbb{Z}
,i\neq j,$ with $x_{i}$ being defined in Proposition 1.

\begin{itemize}
\item Thus, incorporating the above residues expansion into {(\ref{N2}),
making use of the Meijer's }$G$ representation of $\Gamma \left( .,.\right) $
given in \cite[Eq. (06.06.26.0005.01)]{wolfram} as well as performing a
change of variable $\frac{y}{\psi _{i}-\theta _{i}y}+\frac{1}{\theta _{i}}=t$
produces: $\mathcal{B}^{(1,i)}=\mathcal{B}_{\xi _{e}^{2}}^{(1,i)}+\mathcal{B}%
_{\alpha _{e}}^{(1,i)}+\mathcal{B}_{\beta _{e}}^{(1,i)},$ where $\mathcal{B}%
_{x}^{(1,i)}$ is defined in (\ref{B1final}) for $\gamma _{th}<\mathscr{L}%
_{i}\leq 2\gamma _{th}$, with
\begin{align}
\mathcal{S}_{i}\left( y,v\right) & =\frac{\mathscr{L}_{i}}{\theta _{i}}%
b_{y}^{(v)}\Upsilon _{e}^{y}\int_{t=\frac{\gamma _{th}}{\psi _{i}-\theta
_{i}\gamma _{th}}+\frac{1}{\theta _{i}}}^{\infty }t^{-g(y,0,v)}  \notag \\
& \times \frac{G_{1,2}^{2,0}\left( \mathcal{Y}_{e_{i}^{(2)}}^{(ZF)}t\left%
\vert
\begin{array}{c}
-;1 \\
0,-k_{1}-j+p+1;-%
\end{array}%
\right. \right) }{\left( \mathscr{L}_{i}\left( t-\frac{1}{\theta _{i}}%
\right) \right) ^{2-g(y,0,v)}}dt.
\end{align}

Again, using the Meijer's $G$-function definition in \cite[Eq.
(07.34.02.0001.01)]{wolfram} as well as \cite[Eq. (3.194.2)]{integrals} and
through some algebraic manipulations, one obtains:
\begin{align}
\mathcal{S}_{i}\left( y,v\right) & =\mathscr{L}_{i}^{g\left( y,0,v\right)
-1}\Upsilon _{e}^{y}\frac{b_{y}^{(v)}}{\theta _{i}}\frac{1}{2\pi i}%
\int_{C}\left( \frac{\gamma _{th}}{\psi _{i}-\theta _{i}\gamma _{th}}\right)
^{-s-1}  \notag \\
& \times \frac{\Gamma \left( s\right) \Gamma \left( -k_{1}-j+p+1+s\right) }{%
\Gamma \left( 2+s\right) }\left( \mathcal{Y}_{e_{i}^{(2)}}^{(ZF)}\right)
^{-s}  \notag \\
& \times \text{ }_{2}F_{1}\left( s+g\left( y,0,v\right) ,1+s;2+s,1-\frac{%
\mathscr{L}_{i}}{\gamma _{th}}\right) ds.  \label{siint}
\end{align}

with $_{2}F_{1}\left( .,.,.;.\right) $ denotes the Gauss hypergeometric
function \cite[Eqs. (07.23.02.0001.01, 07.23.02.0004.01)]{wolfram}. These
last mentioned identities define this function when the absolute value of
the argument $1-\frac{\mathscr{L}_{i}}{\gamma _{th}}$ is either less or
greater than $1$. For the former case $\left( \text{i.e., }\gamma _{th}<%
\mathscr{L}_{i}\leq 2\gamma _{th}\right) ,$ using the function definition
using eq. (07.23.02.0001.01) of \cite{wolfram}, and based on the Pochhammer
symbol simplification \cite[Eq. (06.10.02.0001.01)]{wolfram} as well as \cite%
[Eq. (07.34.02.0001.01)]{wolfram} , $\mathcal{S}_{i}\left( y,v\right) $
given for the first case of (\ref{B1final}) (i.e., $m=1$) is attained.

Importantly, when $\left( \text{i.e., }\frac{\mathscr{L}_{i}}{\gamma _{th}}%
>2\right) $, and using the second definition of $_{2}F_{1}\left(
.,.,.;z\right) ,$ $z>1$ \cite[Eq. (07.23.02.0004.01)]{wolfram} alongside
with eqs. (06.10.02.0001.01, 07.34.02.0001.01) of \cite{wolfram} and some
algebraic manipulation, $\mathcal{S}_{i}\left( y,v\right) $ in (\ref{siint})
is substituted by the notation $\mathcal{R}_{i}\left( y,v\right) $, where
the resulting expression of $\mathcal{R}_{i}\left( y,v\right) $ are obtained
for the second case of (\ref{B1final}) (i.e., $m=2$).
\end{itemize}

\subsection{Non ZF\ Case}

By using Lemma 1 result alongside with the PDF/CDF of $\gamma
_{e_{i}^{(2)},NZF}$ and $\gamma _{l_{i},NZF}$ in {(\ref{cdfnzf}), we
consider only the case when }$\mathcal{L}^{\left( e_{i}^{(2)}\right)
}<\gamma _{th}<y$ . A{s the PDF\ }$f_{\gamma _{E_{2},NZF}^{(i)}}\left(
z\right) =0$ for $z\geq \mathcal{L}^{\left( e_{i}^{(2)}\right) }$, the
integral definition of $\mathcal{J}_{NZF}\left( y\right) ${\ in Lemma 1 is
positive only for }$z<\mathcal{L}^{\left( e_{i}^{(2)}\right) }$. Thus, in
such an instance, the integral in {(\ref{propip}) becomes}%
\begin{align}
\mathcal{I} &=\underset{\mathcal{K}\left( \gamma _{th}\right) }{\underbrace{%
\left[ \int_{y=\gamma _{th}}^{\infty }f_{\gamma _{1}}\left( y\right)
F_{\gamma _{1}^{(e)}}\left( y\right) dy\right] }}  \notag \\
& \times \underset{\mathcal{J}_{NZF}\left( \mathcal{L}^{\left(
e_{i}^{(2)}\right) }\right) }{\underbrace{\left[ \int_{z=0}^{\mathcal{L}%
^{\left( e_{i}^{(2)}\right) }}f_{\gamma _{e_{i}^{(2)},NZF}}\left( z\right)
F_{\gamma _{l_{i},NZF}}^{c}\left( z\right) dz\right] }};\mathcal{L}^{\left(
e_{i}^{(2)}\right) } <\gamma _{th}.
\end{align}%
Hence, using Lemma 2 result for $y=\mathcal{L}^{\left( e_{i}^{(2)}\right) }$
and Lemma 3 result with $\varphi =\gamma _{th}$ and involving it into {(\ref%
{expI}), one obtains} (\ref{ipclosedform2}).

\section{Appendix F: Proof of Proposition 3}

From {(\ref{propip}), the systems IP\ can be expressed at high SNR }(i.e., $%
\overline{\gamma }_{\Xi }^{(l_{i})},\mu _{1}\rightarrow \infty ,\overline{%
\gamma }_{\Xi }^{(l_{i})}=\epsilon _{\Xi }^{(i)}\mu _{1},\epsilon _{\Xi
}^{(i)}>0)$ as{\ }%
\begin{eqnarray}
P_{int,\Xi }^{\left( i,\infty \right) } &\simeq &1-\int_{y=\gamma
_{th}}^{\infty }f_{\gamma _{1}}^{\infty }\left( y\right) F_{\gamma
_{1}^{(e)}}\left( y\right) dy\int_{z=0}^{y}f_{\gamma _{e_{i}^{(2)},\Xi
}}\left( z\right)  \notag \\
&&\times \left( 1-F_{\gamma _{l_{i},\Xi }}^{\infty }\left( z\right) \right)
dz.  \label{ipass}
\end{eqnarray}

As $\mu _{1}\rightarrow \infty ,$ $\Upsilon \rightarrow 0$ as can be seen
after (\ref{cdfgam1}), which yields that $\Delta _{i}$ given in (\ref{deltai}%
) will be asymptotically represented by considering only least powers of $%
\Upsilon $ in $\mathcal{E}_{i}$'s expression given right after (\ref{deltai}%
), corresponding to $l=0.$\ Therefore

\begin{equation}
\Delta _{i}^{(\infty )}\sim a_{0}^{(i)}\left( \Upsilon z\right) ^{x_{i}},
\end{equation}%
with $x_{i}$ denotes $\xi _{1}^{2},$ $\alpha _{1},$ and $\beta _{1},$ for $%
i=1,2$, and $3$, respectively. Consequently, the CDF\ $F_{\gamma _{1}}(z)$ (%
\ref{cdfgam1}) can be asymptotically approximated using (\ref{cdf11}) as%
\begin{equation}
F_{\gamma _{1}}^{\infty }(z)\sim \left( a_{0}^{(d)}\left( \Upsilon z\right)
^{x_{d}}\mathcal{P}_{1}\right) ^{K},  \label{cdflegas}
\end{equation}%
with $x_{d}=\underset{i=1,2,3}{\min }\left( x_{i}\right) $.

On the other hand, by using the lower-incomplete Gamma expansion \cite[Eq.
(06.06.06.0001.02)]{wolfram} in (\ref{cdfnzf}) and (\ref{cdfgam2}), taking
only the least powers of $\frac{1}{\overline{\gamma }_{\Xi }^{(l_{i})}}$ $($%
i.e., $n_{1}=0),$ one obtains%
\begin{eqnarray}
F_{\gamma _{l_{i},ZF}}^{\infty }\left( z\right) &\simeq &\frac{\lambda
_{l_{i}}v_{l_{i}}z}{\overline{\gamma }_{ZF}^{\left( l_{i}\right) }},
\label{zfasm} \\
F_{\gamma _{l_{i},NZF}}^{\infty }\left( z\right) &\simeq &\frac{\lambda
_{l_{i}}v_{l_{i}}z}{\overline{\gamma }_{NZF}^{(l_{i})}\left( \Psi
^{(l_{i})}-\Theta ^{(l_{i})}z\right) }.  \label{nzfasm}
\end{eqnarray}

\subsection{ZF\ case}

\subsubsection{First Case: $\protect\gamma _{th}>\mathscr{L} _{i}$}

In this case, by making use of integration by parts, (\ref{ipass}) is
expressed as

\begin{align}
P_{int,ZF}^{\left( i,\infty \right) }& =1-\left( \int_{y=\gamma
_{th}}^{\infty }f_{\gamma _{1}}^{\infty }\left( y\right) F_{\gamma
_{1}^{(e)}}\left( y\right) dy\right)  \notag \\
& \times \left( 1-\int_{z=0}^{\mathscr{L}_{i}}f_{\gamma
_{e_{i}^{(2)},ZF}}\left( z\right) F_{\gamma _{l_{i},ZF}}^{\infty }\left(
z\right) dz\right) \\
& =F_{\gamma _{1}}^{\infty }\left( \gamma _{th}\right) F_{\gamma
_{1}^{(e)}}\left( \gamma _{th}\right) +\int_{y=\gamma _{th}}^{\infty
}F_{\gamma _{1}}^{\infty }\left( y\right) f_{\gamma _{1}^{(e)}}\left(
y\right) dy  \notag \\
& +\int_{z=0}^{\mathscr{L}_{i}}f_{\gamma _{e_{i}^{(2)},ZF}}\left( z\right)
F_{\gamma _{l_{i},ZF}}^{\infty }\left( z\right) dz  \notag \\
& -\left( F_{\gamma _{1}}^{\infty }\left( \gamma _{th}\right) F_{\gamma
_{1}^{(e)}}\left( \gamma _{th}\right) +\int_{y=\gamma _{th}}^{\infty
}F_{\gamma _{1}}^{\infty }\left( y\right) f_{\gamma _{1}^{(e)}}\left(
y\right) dy\right)  \notag \\
& \times \int_{z=0}^{\mathscr{L}_{i}}f_{\gamma _{e_{i}^{(2)},ZF}}\left(
z\right) F_{\gamma _{l_{i},ZF}}^{\infty }\left( z\right) dz.
\label{pintasss}
\end{align}

By involving (\ref{cdflegas}) and (\ref{zfasm}) into (\ref{pintasss}), it
can be seen that the diversity order of the first two terms above is $%
G_{d}=Kx_{d},$ $G_{d}=1$ for the third term, and $G_{d}=Kx_{d}+1$ for the
fourth one. Therefore, the IP\ will be expanded by either the first two
terms if $Kx_{d}<1$, or the third term when $Kx_{d}>1.$ Hence, the IP\ is
expressed as%
\begin{equation}
P_{int,ZF}^{\left( i,\infty \right) }\simeq G_{c,ZF}\left( \overline{\gamma }%
_{ZF}^{\left( l_{i}\right) }\right) ^{-G_{d}},\gamma _{th}>\mathscr{L}_{i},
\end{equation}%
with $G_{c,ZF}$ and $G_{d}$ are given in Proposition 3, where $\mathcal{Q}%
^{\left( i,\Xi \right) }=\overline{\gamma }_{\Xi }^{\left( l_{i}\right)
}\int_{z=0}^{\rho }f_{\gamma _{e_{i}^{(2)},\Xi }}\left( z\right) F_{\gamma
_{l_{i},\Xi }}^{\infty }\left( z\right) dz$, with $\rho $ equals $\mathscr{L}%
_{i}$ or $\mathcal{L}^{\left( e_{i}^{(2)}\right) }$ for ZF\ and NZF
scenarios, respectively.

\begin{align}
\mathcal{X}\left( \varphi \right) & =\left( \frac{\overline{\gamma }_{\Xi
}^{\left( l_{i}\right) }}{\epsilon _{\Xi }^{(i)}}\right)
^{Kx_{d}}\int_{y=\varphi }^{\infty }F_{\gamma _{1}}^{\infty }\left( y\right)
f_{\gamma _{1}^{(e)}}\left( y\right) dy  \notag \\
& =\frac{\xi _{e}^{2}}{\Gamma \left( \alpha _{e}\right) \Gamma \left( \beta
_{e}\right) }\left( \frac{a_{0}^{(d)}\xi _{1}^{2}\left( \epsilon _{\Xi
}^{(i)}\Upsilon ^{\prime }\right) ^{x_{d}}}{\Gamma \left( \alpha _{1}\right)
\Gamma \left( \beta _{1}\right) }\right) ^{K}  \notag \\
& \times \left[
\begin{array}{c}
\int_{y=0}^{\infty }z^{Kx_{d}-1}G_{1,3}^{3,0}\left( \Upsilon _{e}z\left\vert
\begin{array}{c}
-;\xi _{e}^{2}+1 \\
\xi _{e}^{2},\alpha _{e},\beta _{e};-%
\end{array}%
\right. \right) \\
-\int_{y=0}^{\varphi }z^{Kx_{d}-1}G_{1,3}^{3,0}\left( \Upsilon
_{e}z\left\vert
\begin{array}{c}
-;\xi _{e}^{2}+1 \\
\xi _{e}^{2},\alpha _{e},\beta _{e};-%
\end{array}%
\right. \right)%
\end{array}%
\right] .
\end{align}

By using the Mellin transform \cite[Eq. (2.9)]{mathai} and the identity \cite%
[Eq. (07.34.21.0003.01)]{wolfram} for $\mathcal{X}\left( \varphi \right) $,
respectively, alongside with some manipulations, one obtains (\ref{Xt}). On
the other hand, by plugging (\ref{zfasm}) and (\ref{cdfeve2}) into $\mathcal{%
Q}^{\left( i,ZF\right) }$ and by using a change of variable $t=\frac{\psi
_{i}}{\psi _{i}-\theta _{i}z}$ with \cite[Eq. (8.350.2)]{integrals}$,$ (\ref%
{qzf}) is reached.

\subsubsection{Second Case: $\protect\gamma _{th}<\mathscr{L} _{i}$}

Likewise, using integration by parts, the\ IP\ in (\ref{ipass}) is expressed
at high SNR\ when $\gamma _{th}<\mathscr{L}_{i}$ as given in (\ref{intermed}%
) at the top of the next page,
\begin{figure*}[t]
{\normalsize 
\setcounter{mytempeqncnt}{\value{equation}}
\setcounter{equation}{99} }%
\begin{eqnarray}
P_{int,ZF}^{\left( i,\infty \right) } &=&1-\int_{y=\gamma _{th}}^{\mathscr{L}%
_{i}}f_{\gamma _{1}}^{\infty }\left( y\right) F_{\gamma _{1}^{(e)}}\left(
y\right) dy\int_{z=0}^{y}f_{\gamma _{e_{i}^{(2)},ZF}}\left( z\right)
F_{\gamma _{l_{i},ZF}}^{c,\infty }\left( z\right) dz  \notag \\
&&-\int_{y=\mathscr{L}_{i}}^{\infty }f_{\gamma _{1}}^{\infty }\left(
y\right) F_{\gamma _{1}^{(e)}}\left( y\right) dy\int_{z=0}^{\mathscr{L}%
_{i}}f_{\gamma _{e_{i}^{(2)},ZF}}\left( z\right) F_{\gamma
_{l_{i},ZF}}^{c,\infty }\left( z\right) dz  \notag \\
&=&\underset{G_{d}=Kx_{d}}{\underbrace{F_{\gamma _{1}}^{\infty }\left( %
\mathscr{L}_{i}\right) F_{\gamma _{1}^{(e)}}\left( \mathscr{L}_{i}\right) }}+%
\underset{G_{d}=Kx_{d}}{\underbrace{\int_{y=\mathscr{L}_{i}}^{\infty
}F_{\gamma _{1}}^{\infty }\left( y\right) f_{\gamma _{1}^{(e)}}\left(
y\right) dy}}+\underset{G_{d}=1}{\underbrace{\int_{z=0}^{\mathscr{L}%
_{i}}f_{\gamma _{e_{i}^{(2)},ZF}}\left( z\right) F_{\gamma
_{l_{i},ZF}}^{\infty }\left( z\right) dz}}  \notag \\
&&-\underset{G_{d}=Kx_{d}+1}{\underbrace{\left( F_{\gamma _{1}}^{\infty
}\left( \mathscr{L}_{i}\right) F_{\gamma _{1}^{(e)}}\left( \mathscr{L}%
_{i}\right) +\int_{y=\mathscr{L}_{i}}^{\infty }F_{\gamma _{1}}^{\infty
}\left( y\right) f_{\gamma _{1}^{(e)}}\left( y\right) dy\right) \int_{z=0}^{%
\mathscr{L}_{i}}f_{\gamma _{e_{i}^{(2)},ZF}}\left( z\right) F_{\gamma
_{l_{i},ZF}}^{\infty }\left( z\right) dz}}  \notag \\
&&-\underset{G_{d}=Kx_{d}+1}{\underbrace{\int_{y=\gamma _{th}}^{\mathscr{L}%
_{i}}f_{\gamma _{1}}^{\infty }\left( y\right) F_{\gamma _{1}^{(e)}}\left(
y\right) dy\int_{z=0}^{y}f_{\gamma _{e_{i}^{(2)},ZF}}\left( z\right)
F_{\gamma _{l_{i},ZF}}^{c,\infty }\left( z\right) dz}}.  \label{intermed}
\end{eqnarray}%
\par
{\normalsize 
\hrulefill 
\vspace*{4pt} }
\end{figure*}

where in a similar way, it can be seen that the IP\ is expanded as given in (%
\ref{gc}), (\ref{aa}) and (\ref{gd}) for $\Xi =ZF.$

\subsection{\protect\bigskip Non-ZF\ case}

In a similar manner to the ZF case, the IP\ can be expanded as given in (\ref%
{pintasss}) by replacing $\mathscr{L}_{i}$ by $\mathcal{L}^{\left(
e_{i}^{(2)}\right) }$ as $P_{int,NZF}^{\left( i,\infty \right) }\simeq
G_{c,NZF}\left( \overline{\gamma }_{NZF}^{\left( l_{i}\right) }\right)
^{-G_{d}}$ for $\gamma _{th}>\mathcal{L}^{\left( e_{i}^{(2)}\right) },$ with
$G_{c,NZF}$ and $G_{d}$ are defined in Proposition 3. By involving (\ref%
{nzfasm}) and (\ref{cdfnzf}) into $\mathcal{Q}^{\left( i,NZF\right) }$
defined in the previous subsection, it yields (\ref{qnzftop}) at the top of
the next page. Finally, by using the change of variable $x=\left\{
\begin{array}{c}
\left( \frac{\mathcal{T}_{i}\Psi ^{(l_{i})}}{\Psi ^{(l_{i})}-\Theta
^{(l_{i})}z}+\Theta ^{\left( e_{i}^{(2)}\right) }\Psi ^{(l_{i})}\right)
^{-1},\text{ }\mathcal{L}^{\left( e_{i}^{(2)}\right) }>\mathcal{L}^{(l_{i})}
\\
\left( -\frac{\mathcal{T}_{i}\Psi ^{\left( e_{i}^{(2)}\right) }}{\Psi
^{\left( e_{i}^{(2)}\right) }-\Theta ^{\left( e_{i}^{(2)}\right) }z}+\Theta
^{(l_{i})}\Psi ^{\left( e_{i}^{(2)}\right) }\right) ^{-1},\text{ }\mathcal{L}%
^{\left( e_{i}^{(2)}\right) }<\mathcal{L}^{(l_{i})}%
\end{array}%
\right. $ with the binomial theorem alongside with \cite[Eqs. (3.352.1),
(3.381.1)]{integrals} for $\mathcal{L}^{\left( e_{i}^{(2)}\right) }>\mathcal{%
L}^{(l_{i})},$ and \cite[Eq. (3.383.4)]{integrals}, \cite[Eq.
(07.45.26.0005.01, 07.34.16.0001.01)]{wolfram} for $\mathcal{L}^{\left(
e_{i}^{(2)}\right) }<\mathcal{L}^{(l_{i})},$ and performing some algebraic
manipulations, one obtains (\ref{qnzf}) given in the proposition.
\begin{figure*}[t]
{\normalsize 
\setcounter{mytempeqncnt}{\value{equation}}
\setcounter{equation}{100} }
\par
\begin{equation}
\mathcal{Q}^{\left( i,NZF\right) }=v_{l_{i}}\Psi ^{\left( e_{i}^{(2)}\right)
}\sum\limits_{n_{2}=0}^{m_{s}^{\left( e_{i}^{(2)}\right) }-1}\frac{\mathcal{U%
}_{i}\left( 0,n_{2}\right) }{\left( \mathcal{Y}_{NZF}^{\left(
e_{i}^{(2)}\right) }\right) ^{-n_{2}-1}}\int_{z=0}^{\mathcal{L}^{\left(
e_{i}^{(2)}\right) }}\frac{z^{n_{2}+1}\exp \left( -\frac{\mathcal{Y}%
_{NZF}^{\left( e_{i}^{(2)}\right) }z}{\left( \Psi ^{\left(
e_{i}^{(2)}\right) }-\Theta ^{\left( e_{i}^{(2)}\right) }z\right) }\right) }{%
\left( \Psi ^{\left( e_{i}^{(2)}\right) }-\Theta ^{\left( e_{i}^{(2)}\right)
}z\right) ^{n_{2}+2}\left( \Psi ^{(l_{i})}-\Theta ^{(l_{i})}z\right) }dz.
\label{qnzftop}
\end{equation}%
\par
{\normalsize 
\hrulefill 
\vspace*{4pt} }
\end{figure*}

\bibliographystyle{IEEEtran}
\bibliography{references}

\end{document}